\documentclass[preprint,journal]{vgtc}            

\onlineid{0}

\ieeedoi{10.1109/TVCG.2024.3372067}

\vgtccategory{Research}

\vgtcpapertype{please specify}

\title{Stepping into the Right Shoes: The Effects of User-Matched Avatar Ethnicity and Gender on Sense of Embodiment in Virtual Reality}

\author{%
  \authororcid{Tiffany D.\ Do}{0000-0003-3323-4586},
  Camille Isabella Protko, and 
  \authororcid{Ryan P.\ McMahan}{0000-0001-9357-9696}
}

\authorfooter{
  \item
  	Tiffany D. Do is with The University of Central Florida.
  	E-mail: tiffany.do@ucf.edu.
  \item
  	Camille Isabella Protko is with The University of Central Florida.
  	E-mail: ca916041@ucf.edu.

  \item Ryan P. McMahan is with The University of Central Florida.
  	E-mail: rpm@ucf.edu.
}

\abstract{%
  In many consumer virtual reality (VR) applications, users embody predefined characters that offer minimal customization options, frequently emphasizing storytelling over user choice. We explore whether matching a user's physical characteristics, specifically ethnicity and gender, with their virtual self-avatar affects their sense of embodiment in VR. We conducted a $2 \times 2$ within-subjects experiment (n=32) with a diverse user population to explore the impact of matching or not matching a user's self-avatar to their ethnicity and gender on their sense of embodiment. Our results indicate that matching the ethnicity of the user and their self-avatar significantly enhances sense of embodiment regardless of gender, extending across various aspects, including appearance, response, and ownership. We also found that matching gender significantly enhanced ownership, suggesting that this aspect is influenced by matching both ethnicity and gender. Interestingly, we found that matching ethnicity specifically affects self-location while matching gender specifically affects one's body ownership.
}


\teaser{
  \centering
  \includegraphics[width=\linewidth]{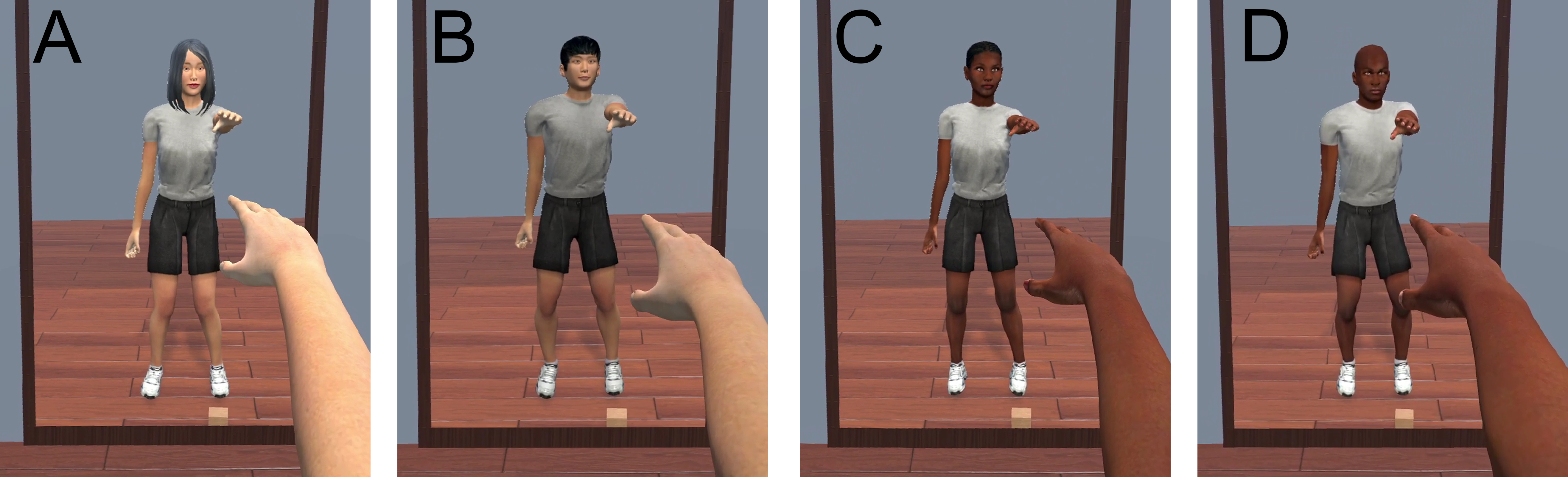}
  \caption{%
  	This example shows an Asian female participant. In our study, we investigated the embodiment effects of matching a self-avatar to the user's ethnicity and gender through four conditions: A) \textit{Complete}: Both ethnicity and gender are matched; B) \textit{Ethnicity}: Ethnicity is matched, but gender is not; C) \textit{Gender}: Gender is matched, but ethnicity is not; D) \textit{None}: Neither ethnicity nor gender are matched.
  }
  \label{fig:teaser}
}

\graphicspath{{figs/}{figures/}{pictures/}{images/}{./}} 

\usepackage{tabu}                      
\usepackage{booktabs}                  
\usepackage{lipsum}                    
\usepackage{mwe}                       

\usepackage{mathptmx}                  

\usepackage{appendix}
\usepackage{amssymb}
\usepackage[normalem]{ulem}
\useunder{\uline}{\ul}{}
\usepackage{cancel}
\newcommand{\add}[1]{{\color{Black}{#1}}}
\newcommand{\rem}[1]{}
\newcommand{\soutpars}[1]{\let\helpcmd\sout\parhelp#1\par\relax\relax}

\begin{document}


\firstsection{Introduction}

\maketitle

In many virtual reality (VR) applications and games, users embody characters that act as their representation within the virtual world. \add{However,} \rem{Nevertheless,}these applications frequently offer limited customization options, prioritizing narrative cohesion over user choice. This constraint becomes apparent in well-known VR titles such as \textit{Half-Life: Alyx}, \textit{Resident Evil Village}, and \textit{Batman: Arkham VR}, where users find themselves restricted to embodying predefined white characters, with no option for gender and ethnicity customization. This raises a fundamental question: does this limitation affect a user's degree of embodiment? 

The study of \rem{avatar embodiment, which describes the effects of self-avatars on its users}\add{sense of embodiment in VR, which describes the combination of sensations experienced while immersed in and controlling a virtual body, \cite{kilteni_sense_2012}}, has garnered increased attention in recent years. We hold a particular interest in this phenomenon because \rem{the degree of} \add{one's sense of }embodiment can play a crucial role in shaping a user's experience within VR, as highlighted by prior research \cite{peck2021}. For example, prior work has shown that the level of embodiment can exert influence over various facets, including distance perception \cite{gonzalez-franco2019} and cognitive performance \cite{juliano_embodiment_2020}. \add{Additionally, an increased sense of embodiment can significantly moderate the outcomes of embodiment illusions, such as the Proteus effect \cite{kilteni2013drumming, malImpact2023}}. However, research investigating similarities between a user and their avatar is sparse \cite{cheymol_beyond_2023}, and there is further limited understanding of how specific factors, such as matched ethnicity and gender, may impact the sense of embodiment.

One of the central challenges in embodiment research is the lack of diversity in avatar representations. For instance, while numerous studies have examined the influence of various factors like avatar appearance and fidelity \cite{latoschik2017, bartl_effects_2022, dewez_influence_2019, schwind2017}, they have frequently used white avatars as the default. This limitation extends to research exploring the similarities between a user and their avatar, including factors like avatar ethnicity and gender. For instance, prior investigations into the effects of embodying avatars of a different gender have predominantly relied on white avatars \cite{lugrin2015, lopez2019, wu2022}. On the other hand, studies exploring the impact of embodying avatars of a different ethnicity \cite{peck2013, kilteni2013drumming, ash2016, salmanowitz_impact_2018} have typically employed gender-matched avatars and limited their participants to mostly individuals of white ethnicity, thus limiting generalizability. Our research goes beyond these prior studies by thoroughly examining the design space of both matched gender and ethnicity to assess how these factors interact and influence \add{one's sense of }embodiment in \rem{virtual reality}\add{VR}.

As VR games and applications continue to gain global traction, it becomes increasingly important to investigate how the similarity between a user's characteristics and their self-avatar can influence their sense of embodiment within the virtual environment. Our research explores the intricate relationship between user ethnicity, gender, and self-avatar in shaping embodiment in VR. We answer the following research questions:
\begin{itemize}
    \itemsep0em 
    \item \textbf{RQ1:} How does matching or not matching a user's ethnicity and their avatar's ethnicity impact \add{sense of }embodiment in VR?
    \item \textbf{RQ2:} How does matching or not matching a user's gender with their avatar's gender impact \add{sense of }embodiment in VR?
    \item \textbf{RQ3:} In which ways do matching gender and ethnicity interact with each \rem{another}\add{other}? 
\end{itemize}

We conducted a $2 \times 2$ within-subjects study involving 32 participants (17 women, 15 men) to investigate the effect of the avatar's ethnicity (matched or not matched) and gender (matched or not matched). In an effort to address the conventional limitations related to participant diversity in embodiment studies, our participant pool encompassed individuals from diverse ethnic backgrounds, including Asian, Black/African-American, Hispanic/Latino, and White populations. Participants engaged in a standard embodiment task \cite{roth2020} using  four avatars: \textit{Complete} (same ethnicity, same gender), \textit{Ethnicity} (same ethnicity, different gender), \textit{Gender} (different ethnicity, same gender), and \textit{None} (different ethnicity, different gender).

\begin{table*}[]\centering
\caption{An overview of avatar embodiment studies investigating the impact of matching user characteristics with their avatar's characteristics.}
\resizebox{\textwidth}{!}{%
\begin{tabular}{lccll|ll|c}
\hline
\textbf{} & \multicolumn{4}{c|}{\textbf{Avatar Factors}} & \multicolumn{2}{c|}{\textbf{\add{Participant Factors}}} & \textbf{Measures} \\ \hline
 & \textbf{Gender} & \textbf{Ethnicity} & \textbf{Avatar Ethnicity} & \textbf{Other} & \textbf{\add{Gender}} & \textbf{\add{Ethnicity}} & \textbf{Sense of Embodiment} \\ \hline
Fribourg et al. \cite{fribourg_avatar_2020} &  &  & Unspecified & Realism, Control & M, F & Unspecified & $\checkmark$ \\
Maselli and Slater  \cite{maselli_building_2013} &  &  & White & Realism & M, F & Unspecified & $\checkmark$ \\
Eubanks et al.  \cite{eubanks_effects_2020} &  &  & Obscured & Tracking level & M & Unspecified & $\checkmark$ \\
Bartl et al. \cite{bartl_effects_2022} &  &  & White & Realism, Clothing & M, F & Unspecified & $\checkmark$ \\
Latoschik et al.  \cite{latoschik2017} &  &  & White & Realism & M, F & Unspecified & $\checkmark$ \\
Dewez et al. \cite{dewez_influence_2019} &  &  & White &  & M, F & Unspecified & $\checkmark$ \\
Fribourg et al. \cite{fribourg_studying_2018} &  &  & Unspecified & Second avatar & M & Unspecified & $\checkmark$ \\
Waltemate et al. \cite{waltemate2018} &  &  & White, Personalized & Personalization, Immersion & M, F & Unspecified & $\checkmark$ \\
Roth and Latoschik  \cite{roth2020} &  &  & White, Personalized & Personalization, Realism & M, F & Unspecified & $\checkmark$ \\
Jo et al.  \cite{jo_impact_2017} &  &  & White, Personalized & Personalization, Quality, Clothing & Unspecified & Unspecified & $\checkmark$ \\
Fiedler et al.  \cite{fiedler_embodiment_2023} &  &  & White, Personalized & Personalization & M, F & Unspecified & $\checkmark$ \\
\add{Dollinger et al. \cite{dollinger2023}} & \multicolumn{1}{l}{} & \multicolumn{1}{l}{} & White, Customized, Personalized & Personalization, Fidelity & M, F & Unspecified & $\checkmark$ \\
\add{Salagean et al. \cite{salagean2023}} & \multicolumn{1}{l}{} & \multicolumn{1}{l}{} & \add{White, Personalized} & \add{Realism} & \add{M, F} & \add{Unspecified} & $\checkmark$ \\
\add{Ries et al. \cite{ries2009}} & \multicolumn{1}{l}{} & \multicolumn{1}{l}{} & \add{Black} & \add{Tracking level, Fidelity} & \add{Unspecified} & \add{Unspecified} & \multicolumn{1}{l}{} \\
\add{Kim et al. \cite{kimBe2023}} & \multicolumn{1}{l}{} & \multicolumn{1}{l}{} & \add{Asian} & \add{Similarity} & M, F & \add{Asian} & $\checkmark$ \\
Radiah et al. \cite{radiah_influence_2023} & $\checkmark$ &  & White, Matched & Personalization & M, F & Unspecified & $\checkmark$ \\
Peck et al.  \cite{peck2020} & $\checkmark$ &  & Matched &  & M, F & 83 White, 42 Other & $\checkmark$ \\
Peck et al. \cite{peck2018} & $\checkmark$ &  & Matched &  & F & White* & $\checkmark$ \\
Lugrin et al.  \cite{lugrin2015} & $\checkmark$ &  & White & Realism & M, F & Unspecified & $\checkmark$ \\
Lopez et al.  \cite{lopez2019} & $\checkmark$ &  & White &  & M & White & $\checkmark$ \\
Wu and Chen   \cite{wu2022} & $\checkmark$ &  & White &  & M, F & Unspecified & $\checkmark$ \\
Schulze et al. \cite{schulze_2019} & $\checkmark$ &  & White &  & M, F & Unspecified &  \\
Chang et al. \cite{chang_stereotype_2019} & $\checkmark$ &  & White &  & M, F & White* &  \\
\add{Schwind et al. \cite{schwind2017}} & $\checkmark$ & \multicolumn{1}{l}{} & \add{White} & \add{Realism} & \add{M, F} & \add{Unspecified} &  \\
Peck et al. \cite{peck2013} &  & $\checkmark$ & Black, White, Alien & Avatar perspective & F & White & $\checkmark$ \\
Kilteni et al. \cite{kilteni2013drumming} &  & $\checkmark$ & Black, White, Obscured & Clothing & M, F & White & $\checkmark$ \\
Marini and Casile \cite{marini_i_2022} &  & $\checkmark$ & Black, White & Mirror presence & F & White & $\checkmark$ \\
Ash \cite{ash2016} &  & $\checkmark$ & Black, White &  & M, F & White* & $\checkmark$ \\
\add{Ambron et al. \cite{ambron2022}} & \multicolumn{1}{l}{} & $\checkmark$ & \add{Black, White, Alien} &  & \add{M, F} & \add{Black, White} & $\checkmark$ \\
Salmanowitz et al. \cite{salmanowitz_impact_2018} &  & $\checkmark$ & Black, White &  & M, F & White &  \\
Chen et al. \cite{chen2021} &  & $\checkmark$ & Asian &  & M, F & Asian &  \\ \hline
\textbf{Ours} & \textbf{$\checkmark$} & \textbf{$\checkmark$} & \textbf{\begin{tabular}[c]{@{}l@{}}Asian, Latino, \\ Black, White\end{tabular}} &  & \textbf{M, F} & \textbf{\begin{tabular}[c]{@{}l@{}}Asian, Latino,\\ Black, White\end{tabular}} & \textbf{$\checkmark$} \\ \hline
\end{tabular}}
*Marked studies reported majority White participants (75\%+), but also included participants of other ethnicities.  
\label{table:related}
\end{table*}

\section{Related Work}
\subsection{Virtual Embodiment}
Virtual embodiment, a key aspect of VR, has gained considerable attention as self-avatars become more popular. Embodiment describes the phenomenon where users are co-located with a virtual avatar and view it from a first-person perspective \cite{peck2021}. Embodiment research delves into how users perceive and interact with their self-avatars \cite{spanlang2014}, and prior research shows that embodiment can impact various aspects of the user experience, such as distance perception \cite{gonzalez-franco2019}, self-enfacement \cite{gonzalez-franco2020}, performance \cite{juliano_embodiment_2020}, and cognitive load \cite{peck2018}. Roth et al. \cite{roth2020} emphasize that embodiment is a part of self-consciousness, primarily driven by the processing of multi-sensory information. We find the degree of embodiment to be important due to its potential to shape the overall user experience in VR. For instance, prior research by Gonzalez-Franco et al. \cite{gonzalez-franco2019} revealed that a higher degree of \add{sense of }embodiment can reduce distance underestimation within VR, while Juliano et al. \cite{juliano_embodiment_2020} found that a higher degree of \add{sense of }embodiment were correlated with increased neurofeedback performance in brain-computer interfaces.

Several scholars have reported that \add{sense of }embodiment comprises various facets. For instance, seminal work by Kilteni et al. \cite{kilteni_sense_2012} describes how embodiment can be broken down into three components: body ownership, defined as ``implications that the body is the source of the experienced sensations" \cite{kilteni_sense_2012}; agency, characterized as ``the feeling of controlling one’s own body movements, and, through them, events in the external environment" \cite{tsakiris2006}; and self-location, representing ``a determinate volume in space where one feels to be located" \cite{kilteni_sense_2012}.

\subsection{Avatar Personalization}
Our research primarily focuses on examining the influence of the resemblance between a user's personal characteristics and those of their self-avatar on the \rem{extent}\add{sense} of embodiment. \add{In recent years, there has been increased interest in the study of how similarity to one's self avatar impacts sense of embodiment, particularly for photorealistic scanned avatars. For example, Kim et al. \cite{kimBe2023} found that a personalized avatar boosted users' sense of embodiment in comparison to a random avatar of the same race and gender.}  Of particular relevance to our study is foundational work on personalization conducted by Waltemate et al. \cite{waltemate2018}\add{ and Salagean et al. \cite{salagean2023}}. \rem{They}\add{Both studies} discovered that a personalized avatar created using photogrammetry elicited a stronger sense of body ownership compared to a generic gender-matched avatar of similar quality. \add{Yet, some studies have reported that personalized avatars may interact with avatar realism. For example, } \rem{A similar study by} Jo et al. \cite{jo_impact_2017} found that participants felt more embodied in a generic avatar with similar clothing to their own in comparison to a realistic point-cloud based representation. \add{Similarly, a recent study by Dollinger et al. \cite{dollinger2023} found that participants embodying a personalized photorealistic avatar had increased sense of embodiment compared to customized and generic avatars, but resulted in higher eeriness and reduced body awareness.} However, these approaches to avatar personalization are not feasible for consumer applications or even some research labs.

Notably, the generic avatars in \add{most of these} previously mentioned studies \add{(\cite{waltemate2018, salagean2023,jo_impact_2017,dollinger2023})}, were depicted as White. This aspect raises important questions about the potential impact of matching or not matching both ethnicity and gender, particularly in the context of generic avatars. For example, would generic avatars of the same ethnicity as the participant induce any differences in \add{sense of }embodiment? Our research expands upon this understanding by specifically investigating how the interaction of matching ethnicity and gender affects embodiment for generic avatars, which have been commonly employed in consumer VR applications and previous embodiment research.

\subsection{Matching Avatar Ethnicity}
Numerous studies have delved into the influence of virtual embodiment on users (e.g., \cite{peck2013, seitz2020, ash2016}). Table \ref{table:related} provides an overview of avatar embodiment research examining matching or not matching avatar ethnicity. Many investigations have centered on how embodying avatars of a different race can impact racial biases (e.g., \cite{peck2013, marini_i_2022}), or influence actions within VR (e.g.,\cite{seitz2020, salmanowitz_impact_2018}). However, our primary focus centers on examining how avatar ethnicity specifically impacts the \textit{degree of the \add{sense of} embodiment}.

Early studies conducted by Peck et al. \cite{peck2013} and Kilteni et al. \cite{kilteni2013drumming}, employing between-groups designs, revealed that White participants who embodied a dark-skinned avatar did not have significant differences in body ownership and \add{sense of }embodiment compared to participants who embodied a light-skinned avatar. \add{Ambron et al. \cite{ambron2022} also found that that both Black and White participants reported comparable ownership levels for light-skinned and dark-skinned hands.} It is important to note that these studies did not employ comprehensive standardized questionnaires to assess embodiment, primarily due to their early nature. Our research advances upon these initial findings by utilizing more extensive and validated questionnaires, including multiple subscales, to precisely investigate the impact of matching race on embodiment. Furthermore, we employ a within-subjects design to assess differences at the individual level.

A more recent study conducted by Marini and Casile \cite{marini_i_2022}, which employed a validated questionnaire \cite{peck2021}, revealed a significant effect of matching the avatar's race on the "appearance" subscale. This subscale includes questions that investigate whether users perceive a connection between their real body and virtual body in terms of appearance. However, it is important to note that Marini and Casile's study exclusively focused on white female users and only provided avatars of white and black ethnicity. This limited scope highlights the direction of our research, as we aim to explore a more diverse participant pool and corresponding avatars.

Freeman and Maloney conducted an interview-based study \cite{freeman_body_2021} and uncovered that users, particularly non-white individuals, regarded the presentation of ethnicity, including skin tone and facial features, as pivotal for self-presentation within VR. Given the importance of ethnicity in shaping the experiences of non-white users, we anticipate that our study, which encompasses participants from various ethnic backgrounds, may yield results that differ from those observed in studies conducted on more homogeneous participant groups (e.g., \cite{peck2013,kilteni2013drumming,marini_i_2022}).

\subsection{Matching Avatar Gender}
Table \ref{table:related} provides an overview of avatar embodiment research examining matching or not matching avatar gender. Of particular relevance to our study is work by Peck et al. \cite{peck2018, peck2020}, which explored the manipulation of avatar gender in a race-matched context across two studies. Although they found that gender matching induced different responses in stereotype lift, it did not yield significant effects for embodiment. It is important to note that Peck et al. \cite{peck2020} acknowledged low reliability in their embodiment questionnaire and recommended the use of more validated measures in future studies. Similarly, in a study investigating how gender can affect implicit bias, Lopez et al. \cite{lopez2019} reported that gender matching did not significantly affect a user's sense of embodiment. However, their study did not employ validated embodiment questionnaires, and instead individually analyzed the responses of five Likert scale questions. Our contribution builds upon these previous investigations by employing robust and validated embodiment questionnaires \cite{peck2021, eubanks2021}, which offer more comprehensive measurements.
 
In contrast to the findings of Lopez et al. \cite{lopez2019} and Peck et al. \cite{peck2018, peck2020}, Lugrin et al. \cite{lugrin2015} reported that individuals who used generic white avatars of the same gender experienced heightened levels of virtual body ownership compared to those who embodied avatars of a different gender during VR fitness training. Similarly, Radiah et al. \cite{radiah_influence_2023} found that participants felt more embodied in a same-gender avatar, but only when they could personalize it. Overall, they reported that participants did not exhibit any significant differences between generic white male and female avatars. However, both studies did not provide information about the ethnicity of the participants, raising questions about whether matching ethnicity may interact with matching gender (i.e., whether a generic avatar that matches the user's ethnicity would have effects of gender). Our study extends this line of research by investigating the combined effects of gender and ethnicity to provide a more comprehensive understanding of their influence on virtual embodiment.

\begin{figure}[h!]
    \centering
    \includegraphics[width=\columnwidth]{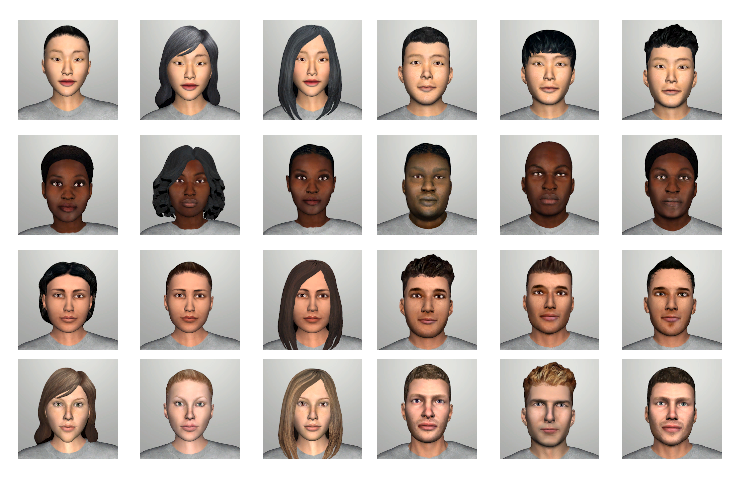}
    \caption{Headshots of the avatars employed in our study, from the Virtual Avatar Library for Inclusion and Diversity (VALID) \cite{do2023valid}, by ethnicity and gender. The top row features Asian avatars, followed by Black/African-American, Hispanic/Latino, and White avatars in subsequent rows. Female avatars are depicted on the left, while male avatars are shown on the right. For each unique ethnicity and gender combination, there were three distinct avatars for participants to choose from. }
    \label{fig:AvatarExamples}
\end{figure}

\section{Methods}
We conducted a $2 \times 2$ within-subjects study to evaluate the effects of matching user ethnicity and gender to their self-avatar. Participants embodied four distinct avatars, which had varying characteristics depending on their self-reported ethnicity and gender: \textit{Complete} (same ethnicity, same gender), \textit{Ethnicity} (same ethnicity, different gender), \textit{Gender} (different ethnicity, same gender), and \textit{None} (different ethnicity, different gender). Participants were able to choose from three avatars for each condition (see Figure \ref{fig:AvatarExamples} for all avatars).

We used avatars from the \textit{Virtual Avatar Library for Inclusion and Diversity} (\textit{VALID}) \cite{do2023valid}, which provides avatars that are statistically validated to be perceived as specific ethnicities and genders (see Figure \ref{fig:AvatarExamples} for examples of the avatars). For ``different gender" avatars, participants embodied an avatar of the opposite gender (man or woman). Our study encompassed participants from diverse ethnic backgrounds, including Asian, Black/African-American, Hispanic/Latino, and White individuals. To ensure that participants perceived different-ethnicity avatars as distinct from their own ethnicity, we selected avatars that were the least representative of the participants' respective ethnicity. This decision was informed by the agreement rates provided by the VALID library. For instance, Asian participants embodied Black avatars as their ``different ethnicity" avatars, as these avatars were perceived as the least representative of Asian ethnicity. For an overview of the selected different-ethnicity avatars corresponding to each participant's ethnicity, please refer to Table \ref{table:DEAvatars}. All avatars wore a gray shirt and black shorts. 

\begin{table}[h!] \centering
\caption{Selected different-ethnicity avatars based on the participant's ethnicity. Agreement rate is the average rate at which the avatar is identified as the participant's ethnicity, as informed by the Virtual Avatar Library for Inclusion and Diversity (VALID). \cite{do2023valid}}
\begin{tabular}{@{}ccc@{}}
\toprule
\textbf{\begin{tabular}[c]{@{}c@{}}Participant \\ Ethnicity\end{tabular}} & \textbf{\begin{tabular}[c]{@{}c@{}}Different-Ethnicity \\ Avatar\end{tabular}} & \textbf{\begin{tabular}[c]{@{}c@{}}Average Agreement \\ Rate\end{tabular}} \\ \midrule
Asian & Black & 0.02 \\
Black & White & \textless{}0.01 \\
Hispanic/Latino & Asian & 0.05 \\
White & Black & \textless{}0.01 \\ \bottomrule
\end{tabular}
\label{table:DEAvatars}
\end{table}

\subsection{Research Hypotheses}
We had the following hypotheses for our research questions:

    \textbf{H1:} Participants will have \rem{a higher degree}\add{an increased sense} of embodiment in an ethnicity-matched avatar, considering prior results indicating that presentation of ethnicity is is important for a user's sense of identity in VR \cite{freeman_body_2021, ambron2022}.
    
    \textbf{H2:} Participants will have \rem{a higher degree}\add{an increased sense} of embodiment in a gender-matched avatar, based on previous findings suggesting that same-gender avatars induce higher levels of \add{sense of }embodiment \cite{lugrin2015}.
    
    \textbf{H3:} Participants will experience \rem{a greater}\add{the greatest} \rem{degree}\add{sense} of embodiment when provided an ethnicity-matched, gender-matched avatar than simply the combination of their individual effects. Due to the lack of prior research in this area, we based this hypothesis on feedback from participants that piloted our study design. 

\subsection{Dependent Variables}

After each condition, we administered two standardized embodiment questionnaires, as described below. We additionally administered an exit survey after all of the conditions to obtain qualitative information from our participants. 

\subsubsection{Standardized Embodiment Questionnaire (SEQ)}
To assess virtual embodiment, we employed the \add{2021} Standardized Embodiment Questionnaire (SEQ) developed by Peck and Gonzalez-Franco \cite{peck2021}. We opted for this questionnaire due to its comprehensive and validated nature, encompassing elements from various previous questionnaires. The SEQ consists of 16 questions, each rated on a 7-point Likert scale. The SEQ generates a main Embodiment score, which is further divided into four subscales: ``Appearance'', ``Response'', ``Ownership'', and ``Multi-Sensory''.

\subsubsection{Preliminary Embodiment Short Questionnaire (pESQ)}
We additionally employed the Preliminary Embodiment Short Questionnaire (pESQ) \cite{eubanks2021} to gain further insights into responses, particularly as it separates  "Ownership" into "Self-Location" and "Body Ownership" and encompasses the three aspects of embodiment as outlined by \cite{kilteni_sense_2012}. \add{We chose to also administer this questionnaire as it is relatively short and measures "Self-Location", which is not measured by other validated questionnaires such as the SEQ \cite{eubanks2021}}. The pESQ consists of five questions, each rated on a 5-point Likert scale. Similar to the SEQ, the pESQ generates an Embodiment score, which is further divided into three subscales: "Agency", "Body Ownership", and "Self-Location".

\subsubsection{Exit Survey}
Upon concluding the experiment, participants participated in an exit survey, allowing them to express their thoughts and insights regarding the avatars. These qualitative responses served as a valuable resource for obtaining deeper context to complement our quantitative findings. Participants were prompted with two key questions: 1) \textit{Which avatar(s) did you feel most embodied by? Why?} and 2) \textit{Which avatar(s) did you feel least embodied by? Why?}

\subsection{Apparatus}
\begin{figure}[h!]
    \centering
    \includegraphics[width=\columnwidth]{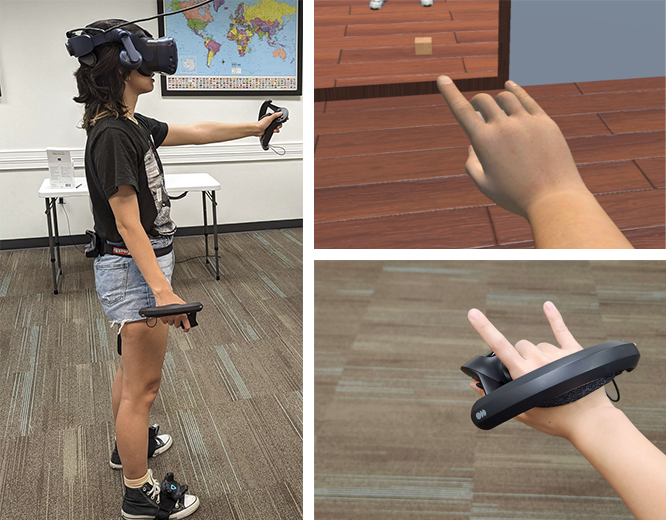}
    \caption{An example of the apparatus. Participants wear a Vive Pro headset, two Valve Index controllers, and three Vive trackers. The Index controllers afford basic finger tracking.}
    \label{fig:apparatus}
\end{figure}

Participants used an HTC Vive Pro and three additional VIVE trackers (2018) on the participants’ feet and lower back. To track their hands, participants used Valve Index controllers, which affords basic finger tracking (see Figure \ref{fig:apparatus} for an example). All movements, including finger tracking, were transferred from these devices to the participants' avatars using FinalIK \footnote{http://root-motion.com/} and SteamVR. 

\begin{figure}[h!] \centering
    \centering
    \includegraphics[width=2.5in]{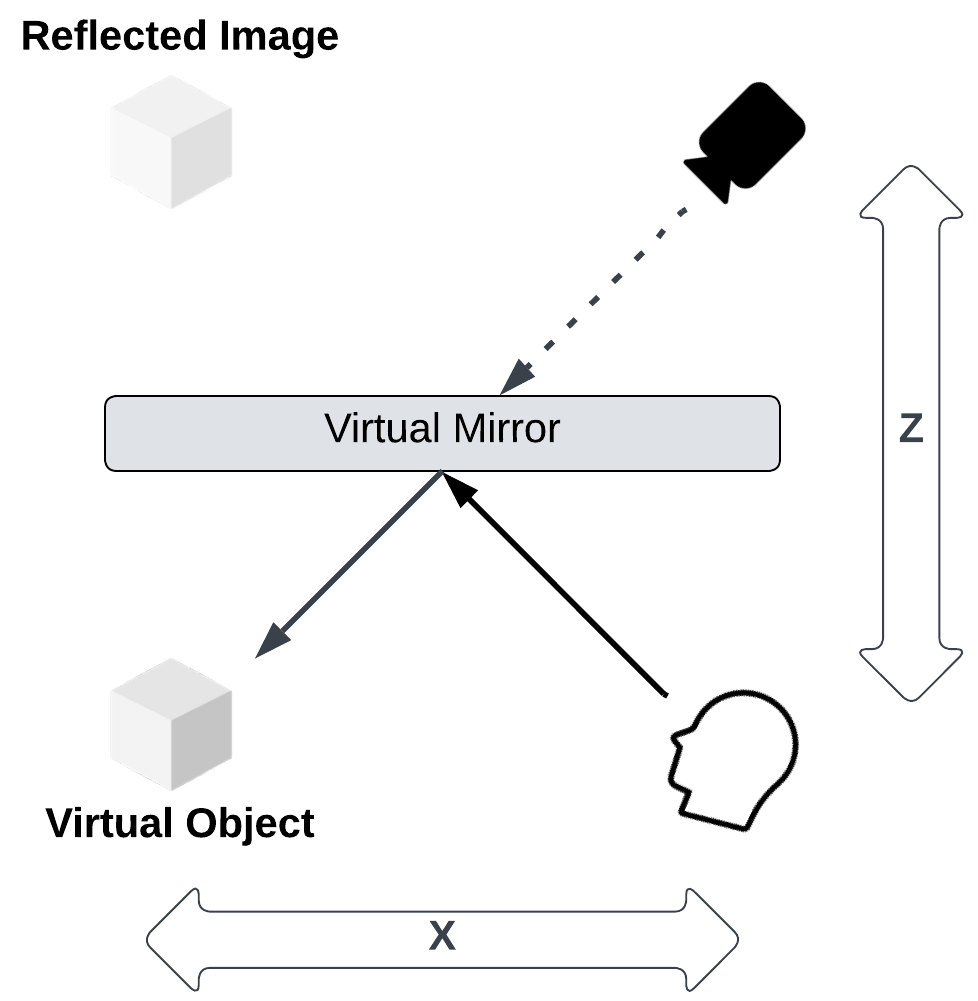}
    \caption{\add{A depiction of our virtual mirror implementation. A secondary virtual camera is positioned across the Z-axis from the user's head position and rotated based on the user's forward direction reflected across the mirror plane to provide the proper reflection in the mirror.}}
    \label{fig:mirror}
\end{figure}

\add{The embodiment simulation was implemented in Unity (2021.3.6f1), featuring a virtual bedroom equipped with a mirror simulation. The virtual environment was illuminated by a soft white directional light (intensity 1.0). The virtual mirror emulated the behavior of a physical mirror by using a secondary virtual camera and a render texture. The secondary virtual camera was positioned on the opposite side of the mirror along the Z-axis, based on the user's head position, and rotated based on the user's forward direction reflected across the mirror plane (see Figure \ref{fig:mirror}). The mirror and wall objects were hidden from this secondary virtual camera to produce the appropriate reflected image within the mirror's render texture.
Users were initially positioned approximately 1.5m away from the mirror.}

\subsection{Procedure}
The study procedure received approval from an Institutional Review Board (IRB) before implementation. The study consisted of one in-person session that lasted approximately 45 minutes. Before arriving to the lab, participants completed a background survey that captured their demographics, ethnicity, gender, education, VR experience, and gaming experience. 

\begin{figure}[h!]
    \centering
    \includegraphics[width=\columnwidth]{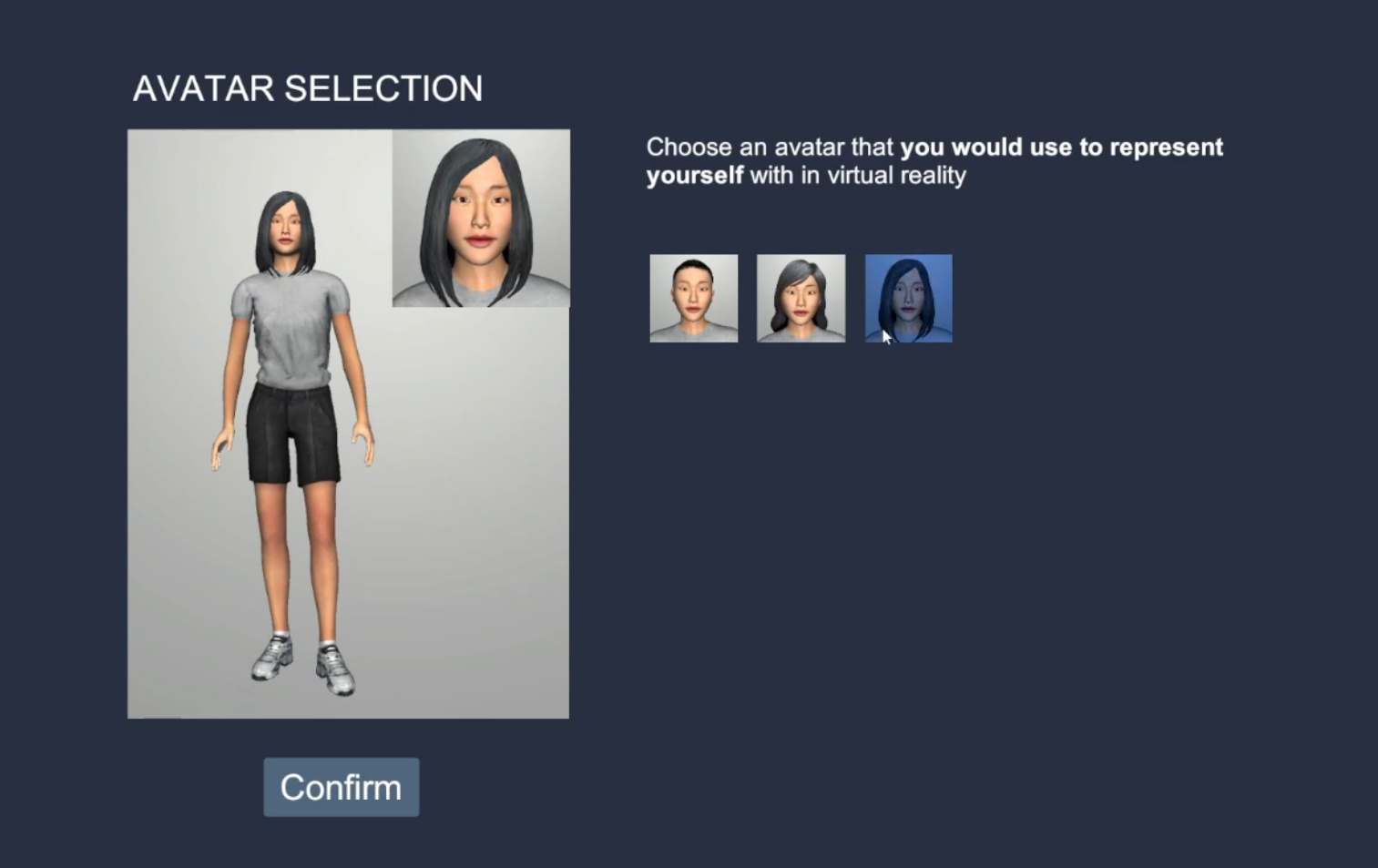}
    \caption{Our avatar selection interface enabled participants to customize their four conditions.}
    \label{fig:selectionInterface}
\end{figure}

Upon arrival to the lab, participants were assigned to one of four Latin squares ordering cohorts for counterbalancing. Subsequently, they were instructed to choose four characters, each corresponding to one of the experimental conditions, using an avatar selection interface displayed on a computer (see Figure \ref{fig:selectionInterface}). The selection screen was dynamically tailored based on each participant's ethnicity and gender, which ensured that participants were presented with avatars that precisely matched the conditions (Complete, Ethnicity, Gender, and None). For example, a participant who identifies as an Asian man would exclusively see avatars representing Asian men when selecting their Complete avatar and avatars representing Black women when choosing their None avatar. Participants were only given the instructions, ``Choose an avatar that you would use to represent yourself in virtual reality" for each selection.

\begin{figure}[h!]
    \centering
    \includegraphics[width=\columnwidth]{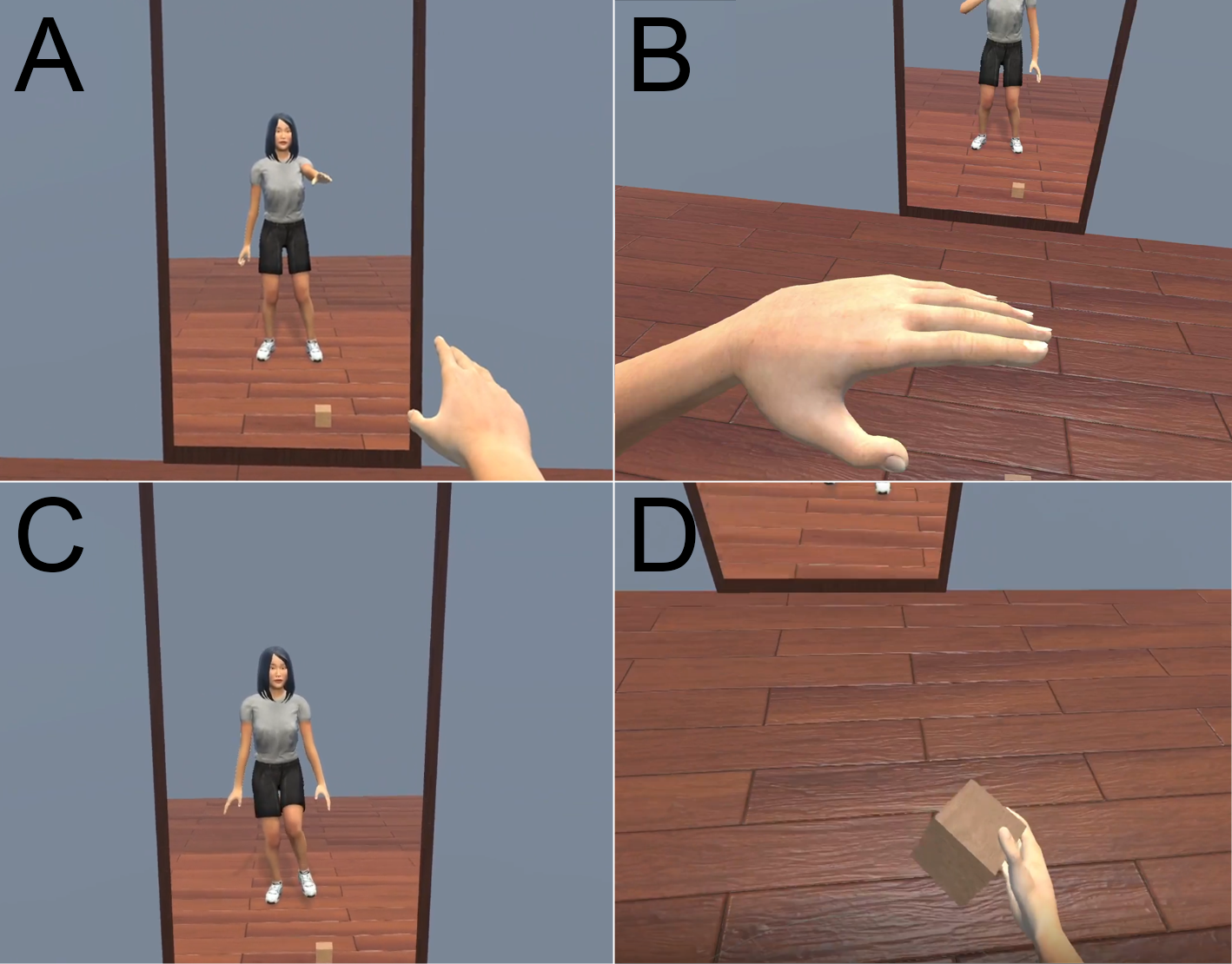}
    \caption{An example of several key tasks from the adapted embodiment procedure. A) Hold your right arm straight forward with your palm facing down. B) Put your left hand with some distance in front of your chest, the palm is facing downwards. C) Take two small steps forward. D) Bend down and place the cube on the floor. }
    \label{fig:tasks}
\end{figure}

Participants proceeded by \add{entering VR and} embodying the first avatar, adhering to the Latin squares ordering assigned to them. They \add{were then instructed to stand straight in order to calibrate} \rem{first calibrated} the avatar's height (using FinalIK's SteamVR calibration), with the avatar remaining concealed from their view until after calibration. \add{The FinalIK calibration system scales the avatar to match the height of the headset.} Subsequently, participants engaged in a standard embodiment procedure, adapted from the protocol developed by Roth and Latoschik \cite{roth2020} (see Figure \ref{fig:tasks}). During these tasks, participants executed actions in front of a virtual mirror, such as ``Hold your right arm forward with your palm facing down" and ``Look at the same hand in the reflection" (see Appendix A). Since our questionnaires encompassed questions concerning feet and object interaction, we included an additional set of six instructions pertaining to feet and object interaction (see Appendix A, Instructions 32-37). \add{After the final task, participants exited VR.}

After each embodiment task, participants filled out the SEQ and pESQ on a computer. This process was then repeated for the remaining three conditions. After the last condition, participants took our short exit survey \add{on a computer}. Participants were then thanked for their time and were compensated with a \$15 USD gift card.   

\subsection{Participants}
A total of 32 participants were recruited from our university via listservs. They provided self-reported demographic information, including gender (17 women and 15 men), ethnicity (7 East Asian, 6 Black/African-American, 11 Hispanic/Latino, and 8 White), and age (M=21.03, SD=3.70). All participants reported normal or corrected vision and indicated the absence of disabilities. Among the participants, 10 reported owning a consumer VR system, while 21 reported playing video games at least weekly.

\begin{figure*}[h!]
    \centering
    \includegraphics[width=\linewidth]{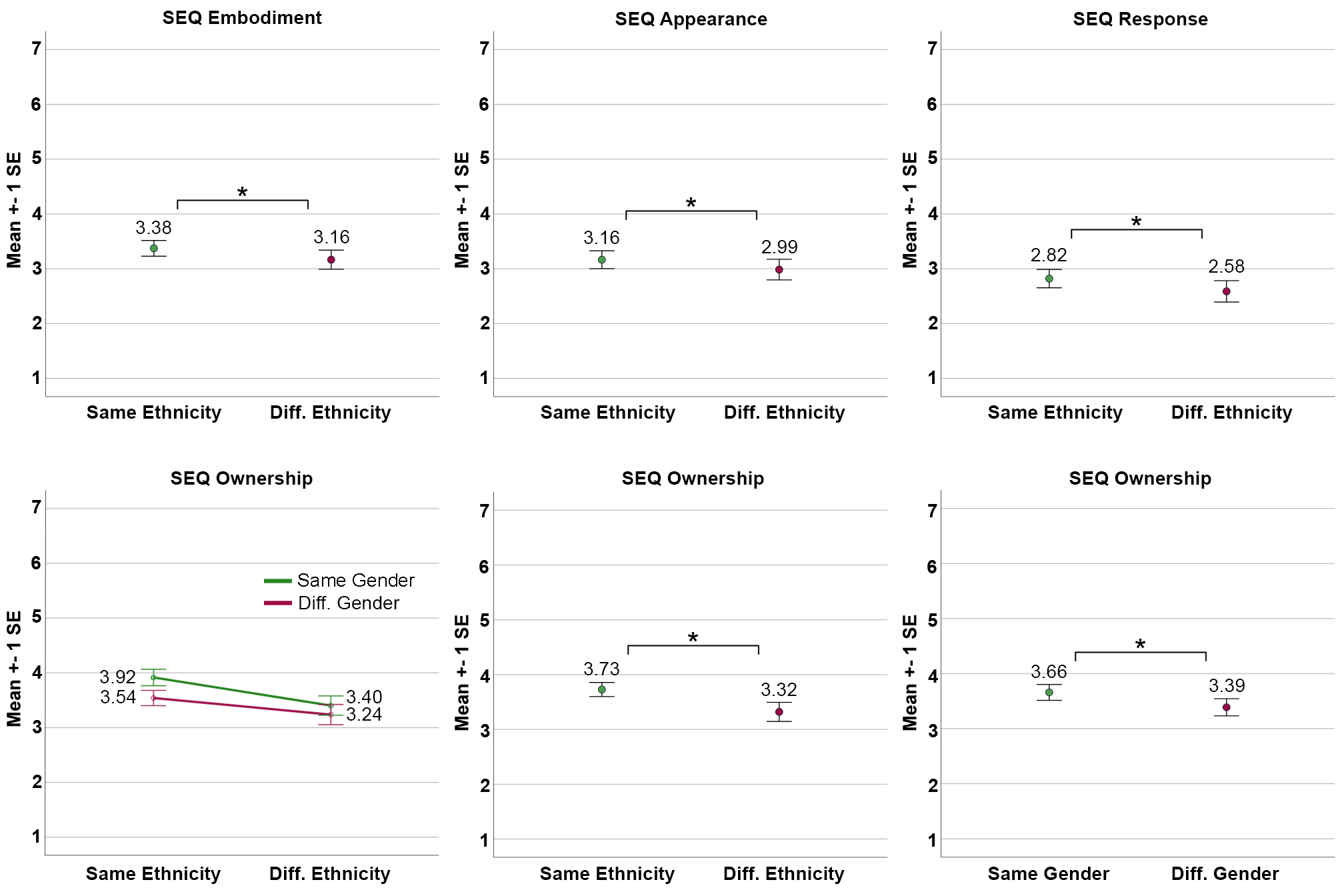}
    \caption{\add{We present mean measurements from the SEQ for all statistically significant results, accompanied by standard error bars (1 SDE). An asterisk (*) denotes significantly different conditions. We display plots grouped by ethnicity conditions for overall Embodiment, Appearance, and Response, given the significant main effects observed for these scales based on ethnicity. For the Ownership scale, which exhibited significant main effects for both ethnicity and gender, we provide plots grouped by both ethnicity and gender. Notably, there were no significant effects identified for the Multi-Sensory subscale.}\rem{Mean measurements from the SEQ for all significant results, presented alongside standard error bars for all experimental conditions. An asterisk (*) indicates significantly different conditions. We show scores for overall Embodiment and its respective subscales: Appearance, Response, and Ownership. There were no significant effects for the Multi-Sensory subscale.}}
    \label{fig:SEQ}
\end{figure*}

\subsection{Data Analysis Approach}
Following the guidelines provided by \cite{peck2021} and \cite{eubanks2021}, we computed the averages of all subscale questions and the total embodiment score. Subsequently, we conducted a Shapiro-Wilk test to determine the normality of the score distributions. Given that this test revealed non-normal distributions for several measures, we opted to employ the Aligned Rank Transform (ART) method \cite{Wobbrock2011} for 2 x 2 factorial analysis of variance. Using this procedure, we tested for interaction and main effects of matched avatar ethnicity and gender.

\section{Results}
The means and standard deviations for all measures can be found in Table \ref{table:results}. We first describe results for the SEQ, and then the pESQ. 

\begin{table}[h!] \centering
\caption{An overview of the means (M) and standard deviations (SD) of standardized questionnaire results. A red asterisk (\textcolor{red}{*}) indicates a significant difference for ethnicity matching, while a red plus (\textcolor{red}{+}) sign indicates a significant difference for gender matching.}
\resizebox{\columnwidth}{!}{%
\begin{tabular}{@{}lcccc@{}}
\toprule
 &
  \textbf{\begin{tabular}[c]{@{}c@{}}\textit{Complete}\\\add{Same Ethnicity} \\ \add{Same Gender} \\ M (SD)\end{tabular}} &
  \textbf{\begin{tabular}[c]{@{}c@{}}\textit{Ethnicity}\\\add{Same Ethnicity} \\ \add{Diff. Gender} \\ M (SD)\end{tabular}} &
  \textbf{\begin{tabular}[c]{@{}c@{}}\textit{Gender}\\\add{Diff. Ethnicity} \\ \add{Same Gender} \\ M (SD)\end{tabular}} &
  \textbf{\begin{tabular}[c]{@{}c@{}}\textit{None}\\\add{Diff. Ethnicity} \\ \add{Diff. Gender} \\ M (SD)\end{tabular}} \\ \midrule
\multicolumn{5}{l}{\textit{\textbf{SEQ}}}                                        \\ \midrule
\textbf{Embodiment\color{red}*}     & 3.48 (0.86) & 3.27 (0.88) & 3.18 (0.97) & 3.15 (1.05) \\
\textbf{Appearance\color{red}*}     & 3.21 (0.94) & 3.11 (1.04) & 2.99 (1.07) & 2.98 (1.11) \\
\textbf{Response\color{red}*}       & 2.92 (0.90) & 2.72 (1.07) & 2.56 (1.04) & 2.62 (1.22) \\
\textbf{Ownership\color{red}*+}     & 3.92 (0.84) & 3.54 (0.78) & 3.40 (0.98) & 3.24 (1.03) \\
\textbf{Multisensory}    & 3.87 (1.07) & 3.69 (0.99) & 3.77 (1.11) & 3.76 (1.19) \\ \midrule
\multicolumn{5}{l}{\textit{\textbf{pESQ}}}                                       \\ \midrule
\textbf{Embodiment}      & 3.74 (0.61) & 3.52 (0.72) & 3.52 (0.72) & 3.40 (0.74) \\
\textbf{Self-Location\color{red}*}  & 3.86 (0.66) & 3.75 (0.88) & 3.48 (1.08) & 3.54 (1.06) \\
\textbf{Agency}          & 4.28 (0.72) & 4.22 (0.58) & 4.28 (0.58) & 4.08 (0.88) \\
\textbf{Body Ownership\color{red}+} & 3.06 (1.07) & 2.59 (1.06) & 2.78 (1.16) & 2.56 (1.08) \\ \bottomrule
\end{tabular}}
\label{table:results}
\end{table}

\subsection{SEQ}
Our analysis of the SEQ data \cite{peck2021} did not reveal any significant interaction effects between matched ethnicity and gender for overall embodiment or any of its subscales. However, we did identify a significant main effect of matched ethnicity on the overall ``Embodiment'' score ($F(1,31) = 7.35$, $p<0.01$, $\eta_{p}^{2}=0.07$). This finding indicates that participants reported higher levels of ``Embodiment" when their ethnicity matched that of their avatar, regardless of matching gender. Conversely, we did not find any main effects of gender matching on the overall ``Embodiment" score.

Upon further examination of the three subscales within the SEQ, we found a significant main effect of matched ethnicity for the ``Appearance" subscale ($F(1,31) = 4.10$, $p = 0.04$, $\eta_{p}^{2}=0.04$), but not for matched gender. Similarly, there was a significant main effect of matched ethnicity for the ``Response" subscale ($F(1,31) = 8.50$, $p<0.001$, $\eta_{p}^{2}=0.08$), but not for matched gender. Hence, participants reported higher levels of ``Appearance" and ``Response" when embodying avatars of the same ethnicity, regardless of the avatar's gender. Interestingly, we also found significant main effects of both matched ethnicity ($F(1,31)=17.87$, $p<0.001$, $\eta_{p}^{2}=0.16$) and matched gender ($F(1,31)=8.31$, $p<0.001$, $\eta_{p}^{2}=0.08$) for the ``Ownership" subscale. 

\begin{figure}[h!]
    \centering
    \includegraphics[width=0.75\columnwidth]{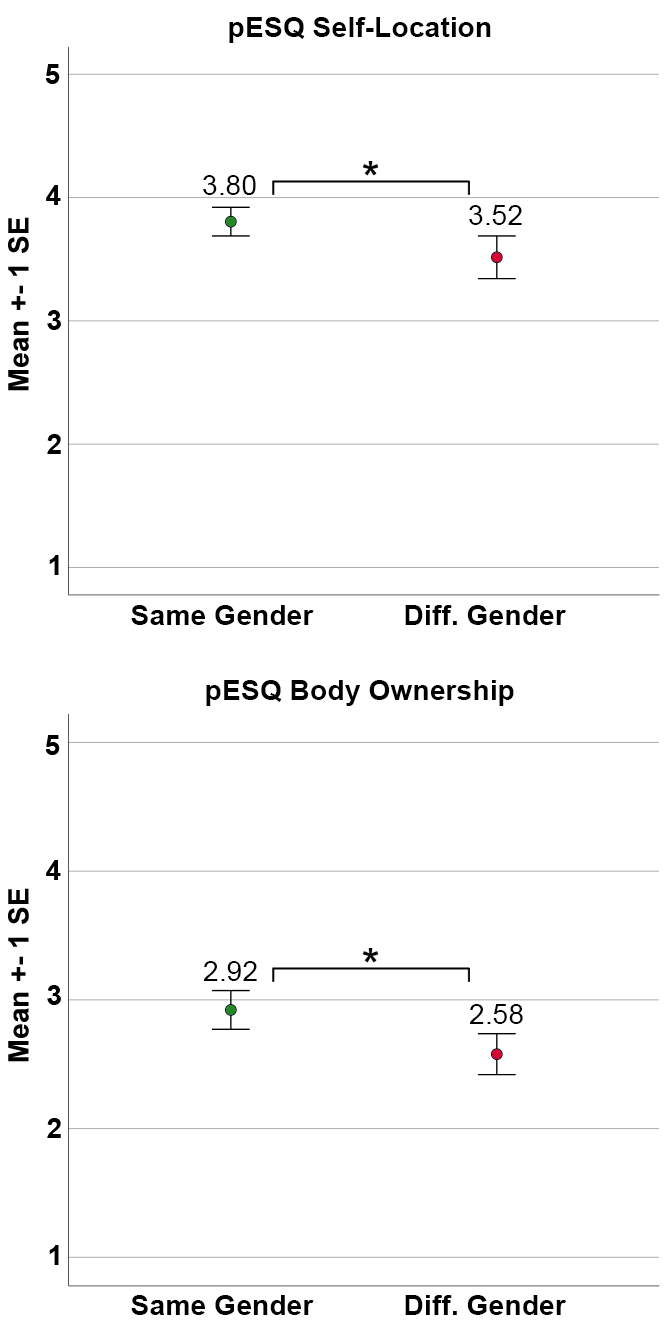}
    \caption{Mean measurements from the pESQ for all significant results, presented alongside standard error bars. An asterisk (*) indicates significantly different conditions. We show scores for Self-location and Body Ownership. There were no significant effects for overall Embodiment or the Agency subscale.}
    \label{fig:pESQ}
\end{figure}

\subsection{pESQ}
Our analysis revealed no significant interaction effects between matched ethnicity and gender for overall embodiment or any subscales of embodiment when using the pESQ \cite{eubanks2021}. We did observe a significant main effect of matched ethnicity on the ``Self-Location" subscale, indicating that participants reported feeling more self-located when their avatar matched their ethnicity ($F(1, 31)=4.74$, $p=0.03$, $\eta_{p}^{2}=0.05$). However, we did not observe a significant main effect of matched gender on the ``Self-Location" subscale ($F(1, 31)=0.03$, $p=0.86$). 

Additionally, there was a significant main effect of matched gender on the ``Body Ownership" subscale, with participants reporting a higher degree of body ownership when the avatar matched their gender ($F(1, 31)=6.06$, $p=0.01$, $\eta_{p}^{2}=0.06$). However, we did not observe a significant main effect of matched ethnicity on the ``Body Ownership" subscale ($F(1, 31)=1.45$, $p=0.23$).

\subsection{Exit Survey}
Additionally, we conducted an exit survey to gather qualitative insights from participants beyond the quantitative measures. Specifically, we inquired about which avatars they felt the most and least embodied by and the reasons for their choices. Table \ref{table:choices} presents the selections made by participants. Participants were allowed to choose multiple avatars in their responses. Notably, most participants felt most embodied in the Complete avatar, and least embodied in the None avatar.  

\begin{table}[h!]
\caption{Responses from the exit survey regarding which avatars participants felt most and least embodied by.}
\begin{tabular}{@{}lcccc@{}}
\toprule
                        & \textbf{Complete} & \textbf{Ethnicity} & \textbf{Gender} & \textbf{None} \\ \midrule
\textbf{Most embodied}  & 27             & 5              & 2              & 1              \\
\textbf{Least embodied} & 1              & 5              & 6              & 25             \\ \bottomrule
\end{tabular}
\label{table:choices}
\end{table}

\section{Discussion}
In this section, we first discuss implications for embodiment research based on our results, and then implications for self-avatar design and development. We conclude the section by discussing the limitations of our work and paths for the future. \add{We found that participants experienced heightened levels of sense of embodiment when they used avatars matching their own ethnicity, irrespective of gender matching. This effect extended across multiple dimensions of embodiment, encompassing aspects pertaining to "appearance", "response", and "ownership" \cite{peck2021}. Furthermore, we observed that gender had a significant effect on feelings of "ownership", suggesting that this aspect of embodiment was influenced by both matching ethnicity and gender.}

\subsection{Matched Avatar Ethnicity Affects Sense of Embodiment}
Our findings suggest that matching a user's ethnicity with their virtual avatar enhances their overall sense of embodiment\rem{ including various facets such as ownership, appearance, and response}. These outcomes contrast with previous research by Peck et al. \cite{peck2013}, Kilteni et al. \cite{kilteni2013drumming}, and Radiah et al. \cite{radiah_influence_2023}, where participants did not report significant differences in \rem{overall}\add{sense of }embodiment when their ethnicity was not matched with their avatar. Several factors may contribute to these differences, including our use of validated questionnaires and a more diverse user population. Whereas these earlier studies exclusively focused on White participants, our research encompassed participants of Asian, Black/African-American, Hispanic/Latino, and White ethnicities, and featured corresponding avatars.

Furthermore, discrepancies in results might also be attributed to the level of tracking involved. In these prior studies, participants were typically seated during the embodiment task and did not engage in interactions involving their feet or lower back. Eubanks et al. \cite{eubanks_effects_2020} have noted that including these body parts in the embodiment experience significantly enhances the sense of embodiment. Consequently, the influence of matched ethnicity on embodiment may become more evident when a broader range of body parts is actively tracked in the virtual experience.

Our findings highlight the significance of ethnicity in \rem{virtual}\add{sense of }embodiment. In the exit survey, a Latino participant noted, \textit{``I felt that the [same-ethnicity avatars] embodied me the most because they resemble physical features of myself...even for the avatar of a different gender, their face looked most closely to mine."} Notably, three participants (2 Asian, 1 Hispanic/Latino) explicitly emphasized that skin tone was the most important factor for their sense of embodiment. These responses echo the sentiments from Freeman and Maloney's interviews \cite{freeman_body_2021}, highlighting the particular significance of ethnic facial features and skin tone for non-white users in VR.

\subsection{Ethnicity and Gender Affect Ownership Differently}
We found that gender significantly affected ownership, as indicated by the SEQ, thus partially supporting H2. Interestingly, the pESQ unveiled further nuanced findings. Eubanks et al. \cite{eubanks2021} note that the SEQ combines self-location and body ownership into a single factor called ``Ownership". However, these factors can be separated to represent ``Body Ownership" and ``Self-Location", respectively. While the SEQ showed that ``Ownership" was affected by both ethnicity and gender matching, further investigation by the pESQ shows that ``Body Ownership" specifically is affected by gender matching, while ``Self-Location" is affected by ethnicity matching. 

Our study revealed that participants experienced a stronger sense of self-location within the virtual environment when their avatar matched their ethnicity, while they reported a greater feeling of body ownership when their avatar matched their gender. These findings align with the insights gathered from our exit surveys, where participants provided comments that reinforced these observations. For instance, one participant remarked, \textit{"I felt like I could picture myself there since [the same-ethnicity avatars] looked similar to me,"} emphasizing the importance of ethnicity in creating a sense of feeling located in the body within the virtual environment. Conversely, another participant stated, \textit{"I felt my body was more true to the proportions that a [same-gender avatar] would have,"} highlighting the role of gender matching in enhancing feelings of body ownership.

\subsection{Implications for VR Research}
Our study holds important implications for the field of VR research. Traditionally, research into the effects of matching or mismatching ethnicity and gender has predominantly focused on white users (e.g., \cite{peck2013, kilteni2013drumming, salmanowitz_impact_2018, lopez2019}) or generic white avatars \cite{bartl_effects_2022, latoschik2017, dewez_influence_2019, lugrin2015, wu2022, chang_stereotype_2019}), which limits the generalizability of findings and may introduce biases into results. For example, requiring an Asian participant to use a White avatar may disrupt their sense of embodiment, potentially leading to unintended consequences in other aspects, such as depth perception \cite{gonzalez-franco2019} or overall cognitive performance \cite{juliano_embodiment_2020}. \add{In previous research by Aseeri and Interrante \cite{aseeri2021} and Herrera and Bailenson \cite{herrera2021}, participants were given the opportunity to choose the skin tone of their self-avatar's hands. We suggest that future studies adopt comparable methodologies, and resources such as the VALID Avatar Library \cite{do2023valid} and Hafnia Hands \cite{pohlHafnia2022} may help with diverse representations.} Additionally, there is a growing recognition that non-white users may place greater importance on the representation of their ethnicity in virtual avatars, considering it a vital aspect of their identity \cite{freeman_body_2021}. As VR continues to gain global popularity, it is imperative for research to embrace more diverse user populations and corresponding avatars to ensure a more comprehensive understanding of virtual embodiment.

Our research shows that matching ethnicity is especially important for overall embodiment. We strongly recommend that VR researchers provide users with the option to customize the ethnicity of their self-avatars, as it can profoundly influence \add{sense of }embodiment and various facets of embodiment. Additionally, we also found that matching avatar gender can also play a role specifically in body ownership. Research in which body ownership is important, such as perspective-taking research, should also allow participants to choose a gender that matches their identity. This customization can enhance the quality of the VR experience and promote a stronger sense of embodiment. We found the Virtual Avatar Library for Inclusion and Diversity (VALID) \cite{do2023valid} very helpful for this purpose.

\subsection{Implications for VR Developers}
Our findings carry important implications for VR application and game developers. We recommend that developers provide players with the option to customize their self-avatar's ethnicity and gender, especially for games that have generic avatars. Striking a balance between narrative cohesion and personalization is crucial for maximizing feelings of embodiment and body ownership. For instance, in games where a character's race or gender is non-essential to the storyline, players should be afforded the ability to customize the main character's characteristics to enhance their sense of connection with the avatar.

Our results found that matching self-avatar ethnicity is especially important for a sense of embodiment, and we thus urge developers to take this into account when designing characters for their applications. While gender matching may only affect body ownership, it may still be important for applications where it is important for users to feel like they "own" their avatar body, such as games with physical combat. This approach can contribute to a more inclusive and engaging VR experience, catering to a wider and more diverse audience.

\subsection{Limitations and Future Work}
While our study included a more diverse range of ethnicities than previous research, it is important to note that our sample size was relatively small ($n = 32$). As a result, we could not explore potential differences in embodiment experiences among various ethnic groups, which might interact with matched avatar ethnicity or gender. In our future research, we plan to recruit a larger participant sample and conduct statistical analyses to explore potential interactions between user ethnicity and matched avatar characteristics. \add{Our participants were also mostly comprised of students from a single university and may not reflect the diversity of the larger more international VR userbase.} Additionally, we only recruited participants identifying as male or female, and we provided only binary avatars for selection. This limitation was partly due to the lack of nonbinary avatars available. However, it is crucial to understand how nonbinary users perceive embodiment in VR and which avatars they feel most connected to.

In our study, we employed generic avatars, which is a common practice in many previous virtual embodiment studies, sourced from the VALID library \cite{do2023valid}. We opted for this library due to its extensive collection of validated avatars that encompass diversity in terms of ethnicity. However, it is important to note that these avatars were not personalized to each individual user, a factor that has been shown to potentially enhance the levels of embodiment \cite{waltemate_impact_2018}. To delve deeper into this aspect, future research should explore how personalization, in conjunction with matched ethnicity and gender, may interplay to influence the phenomenon of virtual embodiment. 

\add{The within-subjects design employed in our study may have been susceptible to demand characteristics \cite{orne2009demand}, where participants discern the study's objectives over time. We utilized Latin Square ordering to counterbalance the sequence of avatars among participants, which may help mitigate some characteristics of demand characteristics \cite{rosenblum1996}. To further validate our findings and address this concern, we conducted a supplementary between-subjects analysis focusing only on participants' initial avatar embodiment. This analysis yielded trends consistent with our within-subjects findings for the SEQ (see Supplementary Materials for additional analyses). However, to ensure the reliability of between-subjects responses, a larger sample size would be necessary for future investigations. Deeper exploration of between-subjects responses with a larger participant pool could validate and refine these outcomes.} 

\add{Furthermore, our decision to include both the SEQ and the pESQ aimed to ensure a comprehensive evaluation of the sense of embodiment, encompassing measurements for 'Self-Location' in addition to the detailed assessments provided by the SEQ. However, employing multiple questionnaires introduces the potential for statistically significant differences to arise purely by chance. We acknowledge that while our results showed significance across both questionnaires, these outcomes might not solely be due to our experimental factors. We suggest that future studies may benefit from employing a single questionnaire to assess the sense of embodiment. This approach could mitigate the likelihood of chance findings due to multiple assessment tools and offer a more focused examination of the experimental variables.}

\subsection{Conclusion}
Our study investigated matching user characteristics, specifically ethnicity and gender, with their self-avatar in VR impacts \add{sense of }embodiment. Through a study involving a diverse group of participants, we gained valuable insights into the effects of matching or mismatching these factors on users' sense of embodiment. We found that matching the ethnicity of the user and their self-avatar significantly enhances overall \add{sense of} embodiment, extending across various aspects, including appearance, response, and ownership. On the other hand, matching gender also significantly affected feelings of ownership, suggesting that this aspect is influenced by both ethnicity and gender congruence. Interestingly, we observed that matching ethnicity significantly affects the self-location aspect of ownership, while matching gender significantly impacts the sense of body ownership.

Our research has important implications for the field of VR embodiment. Historically, VR studies investigating the effects of embodiment have primarily involved limited populations or avatar representations, which may have biased results. We strongly recommend that VR researchers allow users to customize the ethnicity of their self-avatars, as this customization can significantly influence \add{sense of }embodiment. Furthermore, we encourage researchers to embrace more diverse user populations and corresponding avatars to ensure the generalizability and inclusivity of their findings. In summary, our study provides a step forward in understanding the dynamics between user characteristics and their virtual self-avatar in VR environments. As VR technology continues to advance and becomes more globally popular, it is important that research in this field keeps pace by considering the diverse identities of its users.

\section*{Supplemental Materials}
\label{sec:supplemental_materials}All supplemental materials are available on OSF at \url{https://doi.org/10.17605/OSF.IO/R7V9C}, released under a CC BY 4.0 license.
In particular, they include (1) full paper with appendices, (2) full videos of the applications that participants used, (3) additional plots of all data, and (4) additional preliminary analysis of between-subjects data.

\acknowledgments{%
	The authors wish to thank Asma Ahmed for her feedback on early drafts. This work was supported in part by the Doctoral Research Support Award from the College of Graduate Studies at The University of Central Florida.%
}

\bibliographystyle{abbrv-doi-hyperref}

\bibliography{template}

\begin{thebibliography}{10}

\bibitem{ambron2022}
E.~Ambron, S.~Goldstein, A.~Miller, R.~H. Hamilton, and H.~B. Coslett.
\newblock From my skin to your skin: {{Virtual}} image of a hand of different skin color influences movement duration of the real hand in {{Black}} and {{White}} individuals and influences racial bias.
\newblock {\em Frontiers in Virtual Reality}, 3:884189, Dec. 2022. \href{https://doi.org/10.3389/frvir.2022.884189}
{doi: {{%
10\hspace{.1pt}\discretionary{.}{%
}{.}\hspace{.4pt}3389\discretionary{/}{%
}{/}frvir\hspace{.1pt}\discretionary{.}{%
}{.}\hspace{.4pt}2022\hspace{.1pt}\discretionary{.}{%
}{.}\hspace{.4pt}884189}}}


\bibitem{aseeri2021}
S.~Aseeri and V.~Interrante.
\newblock The {{Influence}} of {{Avatar Representation}} on {{Interpersonal Communication}} in {{Virtual Social Environments}}.
\newblock {\em IEEE Transactions on Visualization and Computer Graphics}, 27(5):2608--2617, May 2021. \href{https://doi.org/10.1109/TVCG.2021.3067783}
{doi: {{%
10\hspace{.1pt}\discretionary{.}{%
}{.}\hspace{.4pt}1109\discretionary{/}{%
}{/}TVCG\hspace{.1pt}\discretionary{.}{%
}{.}\hspace{.4pt}2021\hspace{.1pt}\discretionary{.}{%
}{.}\hspace{.4pt}3067783}}}


\bibitem{ash2016}
E.~Ash.
\newblock Priming or {Proteus} {Effect}? {Examining} the {Effects} of \textit{{Avatar} {Race}} on {In}-{Game} {Behavior} and {Post}-{Play} {Aggressive} {Cognition} and {Affect} in {Video} {Games}.
\newblock {\em Games and Culture}, 11(4):422--440, June 2016. \href{https://doi.org/10.1177/1555412014568870}
{doi: {{%
10\hspace{.1pt}\discretionary{.}{%
}{.}\hspace{.4pt}1177\discretionary{/}{%
}{/}1555412014568870}}}


\bibitem{bartl_effects_2022}
A.~Bartl, C.~Merz, D.~Roth, and M.~E. Latoschik.
\newblock The {Effects} of {Avatar} and {Environment} {Design} on {Embodiment}, {Presence}, {Activation}, and {Task} {Load} in a {Virtual} {Reality} {Exercise} {Application}.
\newblock In {\em 2022 {IEEE} {International} {Symposium} on {Mixed} and {Augmented} {Reality} ({ISMAR})}, pp. 260--269. IEEE, Singapore, Singapore, Oct. 2022. \href{https://doi.org/10.1109/ISMAR55827.2022.00041}
{doi: {{%
10\hspace{.1pt}\discretionary{.}{%
}{.}\hspace{.4pt}1109\discretionary{/}{%
}{/}ISMAR55827\hspace{.1pt}\discretionary{.}{%
}{.}\hspace{.4pt}2022\hspace{.1pt}\discretionary{.}{%
}{.}\hspace{.4pt}00041}}}


\bibitem{chang_stereotype_2019}
F.~Chang, M.~Luo, G.~Walton, L.~Aguilar, and J.~Bailenson.
\newblock Stereotype {Threat} in {Virtual} {Learning} {Environments}: {Effects} of {Avatar} {Gender} and {Sexist} {Behavior} on {Women}'s {Math} {Learning} {Outcomes}.
\newblock {\em Cyberpsychology, Behavior, and Social Networking}, 22(10):634--640, Oct. 2019. \href{https://doi.org/10.1089/cyber.2019.0106}
{doi: {{%
10\hspace{.1pt}\discretionary{.}{%
}{.}\hspace{.4pt}1089\discretionary{/}{%
}{/}cyber\hspace{.1pt}\discretionary{.}{%
}{.}\hspace{.4pt}2019\hspace{.1pt}\discretionary{.}{%
}{.}\hspace{.4pt}0106}}}


\bibitem{chen2021}
V.~H.~H. Chen, S.~H.~M. Chan, and Y.~C. Tan.
\newblock Perspective-{Taking} in {Virtual} {Reality} and {Reduction} of {Biases} against {Minorities}.
\newblock {\em Multimodal Technologies and Interaction}, 5(8):42, July 2021. \href{https://doi.org/10.3390/mti5080042}
{doi: {{%
10\hspace{.1pt}\discretionary{.}{%
}{.}\hspace{.4pt}3390\discretionary{/}{%
}{/}mti5080042}}}


\bibitem{cheymol_beyond_2023}
A.~Cheymol, R.~Fribourg, A.~L{\'e}cuyer, J.-M. Normand, and F.~Argelaguet.
\newblock Beyond my {{Real Body}}: {{Characterization}}, {{Impacts}}, {{Applications}} and {{Perspectives}} of ``{{Dissimilar}}'' {{Avatars}} in {{Virtual Reality}}.
\newblock {\em IEEE Transactions on Visualization and Computer Graphics}, 29(11):4426--4437, Nov. 2023. \href{https://doi.org/10.1109/TVCG.2023.3320209}
{doi: {{%
10\hspace{.1pt}\discretionary{.}{%
}{.}\hspace{.4pt}1109\discretionary{/}{%
}{/}TVCG\hspace{.1pt}\discretionary{.}{%
}{.}\hspace{.4pt}2023\hspace{.1pt}\discretionary{.}{%
}{.}\hspace{.4pt}3320209}}}


\bibitem{dewez_influence_2019}
D.~Dewez, R.~Fribourg, F.~Argelaguet, L.~Hoyet, D.~Mestre, M.~Slater, and A.~Lecuyer.
\newblock Influence of {Personality} {Traits} and {Body} {Awareness} on the {Sense} of {Embodiment} in {Virtual} {Reality}.
\newblock In {\em 2019 {IEEE} {International} {Symposium} on {Mixed} and {Augmented} {Reality} ({ISMAR})}, pp. 123--134. IEEE, Beijing, China, Oct. 2019. \href{https://doi.org/10.1109/ISMAR.2019.00-12}
{doi: {{%
10\hspace{.1pt}\discretionary{.}{%
}{.}\hspace{.4pt}1109\discretionary{/}{%
}{/}ISMAR\hspace{.1pt}\discretionary{.}{%
}{.}\hspace{.4pt}2019\hspace{.1pt}\discretionary{.}{%
}{.}\hspace{.4pt}00\discretionary{%
}{-}{-}12}}}


\bibitem{do2023valid}
T.~D. Do, S.~Zelenty, M.~{Gonzalez-Franco}, and R.~P. McMahan.
\newblock {{VALID}}: {{A}} perceptually validated {{Virtual Avatar Library}} for {{Inclusion}} and {{Diversity}}.
\newblock {\em Frontiers in Virtual Reality}. \href{https://doi.org/10.3389/frvir.2023.1248915}
{doi: {{%
10\hspace{.1pt}\discretionary{.}{%
}{.}\hspace{.4pt}3389\discretionary{/}{%
}{/}frvir\hspace{.1pt}\discretionary{.}{%
}{.}\hspace{.4pt}2023\hspace{.1pt}\discretionary{.}{%
}{.}\hspace{.4pt}1248915}}}


\bibitem{dollinger2023}
N.~D{\"o}llinger, M.~Beck, E.~Wolf, D.~Mal, M.~Botsch, M.~E. Latoschik, and C.~Wienrich.
\newblock ``{{If It}}'s {{Not Me It Doesn}}'t {{Make}} a {{Difference}}'' - {{The Impact}} of {{Avatar Personalization}} on user {{Experience}} and {{Body Awareness}} in {{Virtual Reality}}.
\newblock In {\em 2023 {{IEEE International Symposium}} on {{Mixed}} and {{Augmented Reality}} ({{ISMAR}})}, pp. 483--492. {IEEE}, {Sydney, Australia}, Oct. 2023. \href{https://doi.org/10.1109/ISMAR59233.2023.00063}
{doi: {{%
10\hspace{.1pt}\discretionary{.}{%
}{.}\hspace{.4pt}1109\discretionary{/}{%
}{/}ISMAR59233\hspace{.1pt}\discretionary{.}{%
}{.}\hspace{.4pt}2023\hspace{.1pt}\discretionary{.}{%
}{.}\hspace{.4pt}00063}}}


\bibitem{eubanks_effects_2020}
J.~C. Eubanks, A.~G. Moore, P.~A. Fishwick, and R.~P. McMahan.
\newblock The {Effects} of {Body} {Tracking} {Fidelity} on {Embodiment} of an {Inverse}-{Kinematic} {Avatar} for {Male} {Participants}.
\newblock In {\em 2020 {IEEE} {International} {Symposium} on {Mixed} and {Augmented} {Reality} ({ISMAR})}, pp. 54--63. IEEE, Porto de Galinhas, Brazil, Nov. 2020. \href{https://doi.org/10.1109/ISMAR50242.2020.00025}
{doi: {{%
10\hspace{.1pt}\discretionary{.}{%
}{.}\hspace{.4pt}1109\discretionary{/}{%
}{/}ISMAR50242\hspace{.1pt}\discretionary{.}{%
}{.}\hspace{.4pt}2020\hspace{.1pt}\discretionary{.}{%
}{.}\hspace{.4pt}00025}}}


\bibitem{eubanks2021}
J.~C. Eubanks, A.~G. Moore, P.~A. Fishwick, and R.~P. McMahan.
\newblock A {Preliminary} {Embodiment} {Short} {Questionnaire}.
\newblock {\em Frontiers in Virtual Reality}, 2:647896, Apr. 2021. \href{https://doi.org/10.3389/frvir.2021.647896}
{doi: {{%
10\hspace{.1pt}\discretionary{.}{%
}{.}\hspace{.4pt}3389\discretionary{/}{%
}{/}frvir\hspace{.1pt}\discretionary{.}{%
}{.}\hspace{.4pt}2021\hspace{.1pt}\discretionary{.}{%
}{.}\hspace{.4pt}647896}}}


\bibitem{fiedler_embodiment_2023}
M.~L. Fiedler, E.~Wolf, N.~Döllinger, M.~Botsch, M.~E. Latoschik, and C.~Wienrich.
\newblock Embodiment and {Personalization} for {Self}-{Identification} with {Virtual} {Humans}.
\newblock In {\em 2023 {IEEE} {Conference} on {Virtual} {Reality} and {3D} {User} {Interfaces} {Abstracts} and {Workshops} ({VRW})}, pp. 799--800. IEEE, Shanghai, China, Mar. 2023. \href{https://doi.org/10.1109/VRW58643.2023.00242}
{doi: {{%
10\hspace{.1pt}\discretionary{.}{%
}{.}\hspace{.4pt}1109\discretionary{/}{%
}{/}VRW58643\hspace{.1pt}\discretionary{.}{%
}{.}\hspace{.4pt}2023\hspace{.1pt}\discretionary{.}{%
}{.}\hspace{.4pt}00242}}}


\bibitem{freeman_body_2021}
G.~Freeman and D.~Maloney.
\newblock Body, {Avatar}, and {Me}: {The} {Presentation} and {Perception} of {Self} in {Social} {Virtual} {Reality}.
\newblock {\em Proceedings of the ACM on Human-Computer Interaction}, 4(CSCW3):1--27, Jan. 2021. \href{https://doi.org/10.1145/3432938}
{doi: {{%
10\hspace{.1pt}\discretionary{.}{%
}{.}\hspace{.4pt}1145\discretionary{/}{%
}{/}3432938}}}


\bibitem{fribourg_studying_2018}
R.~Fribourg, F.~Argelaguet, L.~Hoyet, and A.~Lecuyer.
\newblock Studying the {Sense} of {Embodiment} in {VR} {Shared} {Experiences}.
\newblock In {\em 2018 {IEEE} {Conference} on {Virtual} {Reality} and {3D} {User} {Interfaces} ({VR})}, pp. 273--280. IEEE, Reutlingen, Mar. 2018. \href{https://doi.org/10.1109/VR.2018.8448293}
{doi: {{%
10\hspace{.1pt}\discretionary{.}{%
}{.}\hspace{.4pt}1109\discretionary{/}{%
}{/}VR\hspace{.1pt}\discretionary{.}{%
}{.}\hspace{.4pt}2018\hspace{.1pt}\discretionary{.}{%
}{.}\hspace{.4pt}8448293}}}


\bibitem{fribourg_avatar_2020}
R.~Fribourg, F.~Argelaguet, A.~Lecuyer, and L.~Hoyet.
\newblock Avatar and {Sense} of {Embodiment}: {Studying} the {Relative} {Preference} {Between} {Appearance}, {Control} and {Point} of {View}.
\newblock {\em IEEE Transactions on Visualization and Computer Graphics}, 26(5):2062--2072, May 2020. \href{https://doi.org/10.1109/TVCG.2020.2973077}
{doi: {{%
10\hspace{.1pt}\discretionary{.}{%
}{.}\hspace{.4pt}1109\discretionary{/}{%
}{/}TVCG\hspace{.1pt}\discretionary{.}{%
}{.}\hspace{.4pt}2020\hspace{.1pt}\discretionary{.}{%
}{.}\hspace{.4pt}2973077}}}


\bibitem{gonzalez-franco2019}
M.~Gonzalez-Franco, P.~Abtahi, and A.~Steed.
\newblock Individual {Differences} in {Embodied} {Distance} {Estimation} in {Virtual} {Reality}.
\newblock In {\em 2019 {IEEE} {Conference} on {Virtual} {Reality} and {3D} {User} {Interfaces} ({VR})}, pp. 941--943. IEEE, Osaka, Japan, Mar. 2019. \href{https://doi.org/10.1109/VR.2019.8798348}
{doi: {{%
10\hspace{.1pt}\discretionary{.}{%
}{.}\hspace{.4pt}1109\discretionary{/}{%
}{/}VR\hspace{.1pt}\discretionary{.}{%
}{.}\hspace{.4pt}2019\hspace{.1pt}\discretionary{.}{%
}{.}\hspace{.4pt}8798348}}}


\bibitem{gonzalez-franco2020}
M.~Gonzalez-Franco, A.~Steed, S.~Hoogendyk, and E.~Ofek.
\newblock Using {Facial} {Animation} to {Increase} the {Enfacement} {Illusion} and {Avatar} {Self}-{Identification}.
\newblock {\em IEEE Transactions on Visualization and Computer Graphics}, 26(5):2023--2029, May 2020. \href{https://doi.org/10.1109/TVCG.2020.2973075}
{doi: {{%
10\hspace{.1pt}\discretionary{.}{%
}{.}\hspace{.4pt}1109\discretionary{/}{%
}{/}TVCG\hspace{.1pt}\discretionary{.}{%
}{.}\hspace{.4pt}2020\hspace{.1pt}\discretionary{.}{%
}{.}\hspace{.4pt}2973075}}}


\bibitem{herrera2021}
F.~Herrera and J.~N. Bailenson.
\newblock Virtual reality perspective-taking at scale: {{Effect}} of avatar representation, choice, and head movement on prosocial behaviors.
\newblock {\em New Media \& Society}, 23(8):2189--2209, Aug. 2021. \href{https://doi.org/10.1177/1461444821993121}
{doi: {{%
10\hspace{.1pt}\discretionary{.}{%
}{.}\hspace{.4pt}1177\discretionary{/}{%
}{/}1461444821993121}}}


\bibitem{jo_impact_2017}
D.~Jo, K.~Kim, G.~F. Welch, W.~Jeon, Y.~Kim, K.-H. Kim, and G.~J. Kim.
\newblock The impact of avatar-owner visual similarity on body ownership in immersive virtual reality.
\newblock In {\em Proceedings of the 23rd {ACM} {Symposium} on {Virtual} {Reality} {Software} and {Technology}}, pp. 1--2. ACM, Gothenburg Sweden, Nov. 2017. \href{https://doi.org/10.1145/3139131.3141214}
{doi: {{%
10\hspace{.1pt}\discretionary{.}{%
}{.}\hspace{.4pt}1145\discretionary{/}{%
}{/}3139131\hspace{.1pt}\discretionary{.}{%
}{.}\hspace{.4pt}3141214}}}


\bibitem{juliano_embodiment_2020}
J.~M. Juliano, R.~P. Spicer, A.~Vourvopoulos, S.~Lefebvre, K.~Jann, T.~Ard, E.~Santarnecchi, D.~M. Krum, and S.-L. Liew.
\newblock Embodiment {Is} {Related} to {Better} {Performance} on a {Brain}–{Computer} {Interface} in {Immersive} {Virtual} {Reality}: {A} {Pilot} {Study}.
\newblock {\em Sensors}, 20(4):1204, Feb. 2020. \href{https://doi.org/10.3390/s20041204}
{doi: {{%
10\hspace{.1pt}\discretionary{.}{%
}{.}\hspace{.4pt}3390\discretionary{/}{%
}{/}s20041204}}}


\bibitem{kilteni2013drumming}
K.~Kilteni, I.~Bergstrom, and M.~Slater.
\newblock Drumming in immersive virtual reality: the body shapes the way we play.
\newblock {\em IEEE transactions on visualization and computer graphics}, 19(4):597--605, 2013.

\bibitem{kilteni_sense_2012}
K.~Kilteni, R.~Groten, and M.~Slater.
\newblock The {Sense} of {Embodiment} in {Virtual} {Reality}.
\newblock {\em Presence: Teleoperators and Virtual Environments}, 21(4):373--387, Nov. 2012. \href{https://doi.org/10.1162/PRES_a_00124}
{doi: {{%
10\hspace{.1pt}\discretionary{.}{%
}{.}\hspace{.4pt}1162\discretionary{/}{%
}{/}PRES\_a\_00124}}}


\bibitem{kimBe2023}
H.~Kim, J.~Park, and I.-K. Lee.
\newblock ``{{To}} be or {{Not}} to be {{Me}}?'': {{Exploration}} of {{Self-Similar Effects}} of {{Avatars}} on {{Social Virtual Reality Experiences}}.
\newblock {\em IEEE Transactions on Visualization and Computer Graphics}, 29(11):4794--4804, Nov. 2023. \href{https://doi.org/10.1109/TVCG.2023.3320240}
{doi: {{%
10\hspace{.1pt}\discretionary{.}{%
}{.}\hspace{.4pt}1109\discretionary{/}{%
}{/}TVCG\hspace{.1pt}\discretionary{.}{%
}{.}\hspace{.4pt}2023\hspace{.1pt}\discretionary{.}{%
}{.}\hspace{.4pt}3320240}}}


\bibitem{latoschik2017}
M.~E. Latoschik, D.~Roth, D.~Gall, J.~Achenbach, T.~Waltemate, and M.~Botsch.
\newblock The effect of avatar realism in immersive social virtual realities.
\newblock In {\em Proceedings of the 23rd {ACM} {Symposium} on {Virtual} {Reality} {Software} and {Technology}}, pp. 1--10. ACM, Gothenburg Sweden, Nov. 2017. \href{https://doi.org/10.1145/3139131.3139156}
{doi: {{%
10\hspace{.1pt}\discretionary{.}{%
}{.}\hspace{.4pt}1145\discretionary{/}{%
}{/}3139131\hspace{.1pt}\discretionary{.}{%
}{.}\hspace{.4pt}3139156}}}


\bibitem{lopez2019}
S.~Lopez, Y.~Yang, K.~Beltran, S.~J. Kim, J.~Cruz~Hernandez, C.~Simran, B.~Yang, and B.~F. Yuksel.
\newblock Investigating implicit gender bias and embodiment of white males in virtual reality with full body visuomotor synchrony.
\newblock In {\em Proceedings of the 2019 CHI Conference on Human Factors in Computing Systems}, CHI '19, p. 1–12. Association for Computing Machinery, New York, NY, USA, 2019. \href{https://doi.org/10.1145/3290605.3300787}
{doi: {{%
10\hspace{.1pt}\discretionary{.}{%
}{.}\hspace{.4pt}1145\discretionary{/}{%
}{/}3290605\hspace{.1pt}\discretionary{.}{%
}{.}\hspace{.4pt}3300787}}}


\bibitem{lugrin2015}
J.-L. Lugrin, M.~Landeck, and M.~E. Latoschik.
\newblock Avatar embodiment realism and virtual fitness training.
\newblock In {\em 2015 {IEEE} {Virtual} {Reality} ({VR})}, pp. 225--226. IEEE, Arles, Camargue, Provence, France, Mar. 2015. \href{https://doi.org/10.1109/VR.2015.7223377}
{doi: {{%
10\hspace{.1pt}\discretionary{.}{%
}{.}\hspace{.4pt}1109\discretionary{/}{%
}{/}VR\hspace{.1pt}\discretionary{.}{%
}{.}\hspace{.4pt}2015\hspace{.1pt}\discretionary{.}{%
}{.}\hspace{.4pt}7223377}}}


\bibitem{malImpact2023}
D.~Mal, E.~Wolf, N.~D{\"o}llinger, C.~Wienrich, and M.~E. Latoschik.
\newblock The {{Impact}} of {{Avatar}} and {{Environment Congruence}} on {{Plausibility}}, {{Embodiment}}, {{Presence}}, and the {{Proteus Effect}} in {{Virtual Reality}}.
\newblock {\em IEEE Transactions on Visualization and Computer Graphics}, 29(5):2358--2368, May 2023. \href{https://doi.org/10.1109/TVCG.2023.3247089}
{doi: {{%
10\hspace{.1pt}\discretionary{.}{%
}{.}\hspace{.4pt}1109\discretionary{/}{%
}{/}TVCG\hspace{.1pt}\discretionary{.}{%
}{.}\hspace{.4pt}2023\hspace{.1pt}\discretionary{.}{%
}{.}\hspace{.4pt}3247089}}}


\bibitem{marini_i_2022}
M.~Marini and A.~Casile.
\newblock I can see my virtual body in a mirror: {The} role of visual perspective in changing implicit racial attitudes using virtual reality.
\newblock {\em Frontiers in Psychology}, 13:989582, Nov. 2022. \href{https://doi.org/10.3389/fpsyg.2022.989582}
{doi: {{%
10\hspace{.1pt}\discretionary{.}{%
}{.}\hspace{.4pt}3389\discretionary{/}{%
}{/}fpsyg\hspace{.1pt}\discretionary{.}{%
}{.}\hspace{.4pt}2022\hspace{.1pt}\discretionary{.}{%
}{.}\hspace{.4pt}989582}}}


\bibitem{maselli_building_2013}
A.~Maselli and M.~Slater.
\newblock The building blocks of the full body ownership illusion.
\newblock {\em Frontiers in Human Neuroscience}, 7, 2013. \href{https://doi.org/10.3389/fnhum.2013.00083}
{doi: {{%
10\hspace{.1pt}\discretionary{.}{%
}{.}\hspace{.4pt}3389\discretionary{/}{%
}{/}fnhum\hspace{.1pt}\discretionary{.}{%
}{.}\hspace{.4pt}2013\hspace{.1pt}\discretionary{.}{%
}{.}\hspace{.4pt}00083}}}


\bibitem{orne2009demand}
M.~T. Orne.
\newblock Demand characteristics and the concept of quasi-controls.
\newblock {\em Artifacts in behavioral research}, 110:110--137, 2009.

\bibitem{peck2018}
T.~C. Peck, M.~Doan, K.~A. Bourne, and J.~J. Good.
\newblock The {Effect} of {Gender} {Body}-{Swap} {Illusions} on {Working} {Memory} and {Stereotype} {Threat}.
\newblock {\em IEEE Transactions on Visualization and Computer Graphics}, 24(4):1604--1612, Apr. 2018.
\newblock Conference Name: IEEE Transactions on Visualization and Computer Graphics. \href{https://doi.org/10.1109/TVCG.2018.2793598}
{doi: {{%
10\hspace{.1pt}\discretionary{.}{%
}{.}\hspace{.4pt}1109\discretionary{/}{%
}{/}TVCG\hspace{.1pt}\discretionary{.}{%
}{.}\hspace{.4pt}2018\hspace{.1pt}\discretionary{.}{%
}{.}\hspace{.4pt}2793598}}}


\bibitem{peck2021}
T.~C. Peck and M.~Gonzalez-Franco.
\newblock Avatar {Embodiment}. {A} {Standardized} {Questionnaire}.
\newblock {\em Frontiers in Virtual Reality}, 1:575943, Feb. 2021. \href{https://doi.org/10.3389/frvir.2020.575943}
{doi: {{%
10\hspace{.1pt}\discretionary{.}{%
}{.}\hspace{.4pt}3389\discretionary{/}{%
}{/}frvir\hspace{.1pt}\discretionary{.}{%
}{.}\hspace{.4pt}2020\hspace{.1pt}\discretionary{.}{%
}{.}\hspace{.4pt}575943}}}


\bibitem{peck2020}
T.~C. Peck, J.~J. Good, and K.~A. Bourne.
\newblock Inducing and {Mitigating} {Stereotype} {Threat} {Through} {Gendered} {Virtual} {Body}-{Swap} {Illusions}.
\newblock In {\em Proceedings of the 2020 {CHI} {Conference} on {Human} {Factors} in {Computing} {Systems}}, pp. 1--13. ACM, Honolulu HI USA, Apr. 2020. \href{https://doi.org/10.1145/3313831.3376419}
{doi: {{%
10\hspace{.1pt}\discretionary{.}{%
}{.}\hspace{.4pt}1145\discretionary{/}{%
}{/}3313831\hspace{.1pt}\discretionary{.}{%
}{.}\hspace{.4pt}3376419}}}


\bibitem{peck2013}
T.~C. Peck, S.~Seinfeld, S.~M. Aglioti, and M.~Slater.
\newblock Putting yourself in the skin of a black avatar reduces implicit racial bias.
\newblock {\em Consciousness and Cognition}, 22(3):779--787, Sept. 2013. \href{https://doi.org/10.1016/j.concog.2013.04.016}
{doi: {{%
10\hspace{.1pt}\discretionary{.}{%
}{.}\hspace{.4pt}1016\discretionary{/}{%
}{/}j\hspace{.1pt}\discretionary{.}{%
}{.}\hspace{.4pt}concog\hspace{.1pt}\discretionary{.}{%
}{.}\hspace{.4pt}2013\hspace{.1pt}\discretionary{.}{%
}{.}\hspace{.4pt}04\hspace{.1pt}\discretionary{.}{%
}{.}\hspace{.4pt}016}}}


\bibitem{pohlHafnia2022}
H.~Pohl and A.~Mottelson.
\newblock Hafnia {{Hands}}: {{A Multi-Skin Hand Texture Resource}} for {{Virtual Reality Research}}.
\newblock {\em Frontiers in Virtual Reality}, 3:719506, May 2022. \href{https://doi.org/10.3389/frvir.2022.719506}
{doi: {{%
10\hspace{.1pt}\discretionary{.}{%
}{.}\hspace{.4pt}3389\discretionary{/}{%
}{/}frvir\hspace{.1pt}\discretionary{.}{%
}{.}\hspace{.4pt}2022\hspace{.1pt}\discretionary{.}{%
}{.}\hspace{.4pt}719506}}}


\bibitem{radiah_influence_2023}
R.~Radiah, D.~Roth, F.~Alt, and Y.~Abdelrahman.
\newblock The {Influence} of {Avatar} {Personalization} on {Emotions} in {VR}.
\newblock {\em Multimodal Technologies and Interaction}, 7(4):38, Mar. 2023. \href{https://doi.org/10.3390/mti7040038}
{doi: {{%
10\hspace{.1pt}\discretionary{.}{%
}{.}\hspace{.4pt}3390\discretionary{/}{%
}{/}mti7040038}}}


\bibitem{ries2009}
B.~Ries, V.~Interrante, M.~Kaeding, and L.~Phillips.
\newblock Analyzing the effect of a virtual avatar's geometric and motion fidelity on ego-centric spatial perception in immersive virtual environments.
\newblock In {\em Proceedings of the 16th {{ACM Symposium}} on {{Virtual Reality Software}} and {{Technology}}}, pp. 59--66. {ACM}, {Kyoto Japan}, Nov. 2009. \href{https://doi.org/10.1145/1643928.1643943}
{doi: {{%
10\hspace{.1pt}\discretionary{.}{%
}{.}\hspace{.4pt}1145\discretionary{/}{%
}{/}1643928\hspace{.1pt}\discretionary{.}{%
}{.}\hspace{.4pt}1643943}}}


\bibitem{rosenblum1996}
L.~D. Rosenblum and H.~M. Salda{\~n}a.
\newblock An audiovisual test of kinematic primitives for visual speech perception.
\newblock {\em Journal of Experimental Psychology: Human Perception and Performance}, 22(2):318--331, 1996. \href{https://doi.org/10.1037/0096-1523.22.2.318}
{doi: {{%
10\hspace{.1pt}\discretionary{.}{%
}{.}\hspace{.4pt}1037\discretionary{/}{%
}{/}0096\discretionary{%
}{-}{-}1523\hspace{.1pt}\discretionary{.}{%
}{.}\hspace{.4pt}22\hspace{.1pt}\discretionary{.}{%
}{.}\hspace{.4pt}2\hspace{.1pt}\discretionary{.}{%
}{.}\hspace{.4pt}318}}}


\bibitem{roth2020}
D.~Roth and M.~E. Latoschik.
\newblock Construction of the {Virtual} {Embodiment} {Questionnaire} ({VEQ}).
\newblock {\em IEEE Transactions on Visualization and Computer Graphics}, 26(12):3546--3556, Dec. 2020. \href{https://doi.org/10.1109/TVCG.2020.3023603}
{doi: {{%
10\hspace{.1pt}\discretionary{.}{%
}{.}\hspace{.4pt}1109\discretionary{/}{%
}{/}TVCG\hspace{.1pt}\discretionary{.}{%
}{.}\hspace{.4pt}2020\hspace{.1pt}\discretionary{.}{%
}{.}\hspace{.4pt}3023603}}}


\bibitem{salagean2023}
A.~Salagean, E.~Crellin, M.~Parsons, D.~Cosker, and D.~Stanton~Fraser.
\newblock Meeting {{Your Virtual Twin}}: {{Effects}} of {{Photorealism}} and {{Personalization}} on {{Embodiment}}, {{Self-Identification}} and {{Perception}} of {{Self-Avatars}} in {{Virtual Reality}}.
\newblock In {\em Proceedings of the 2023 {{CHI Conference}} on {{Human Factors}} in {{Computing Systems}}}, pp. 1--16. {ACM}, {Hamburg Germany}, Apr. 2023. \href{https://doi.org/10.1145/3544548.3581182}
{doi: {{%
10\hspace{.1pt}\discretionary{.}{%
}{.}\hspace{.4pt}1145\discretionary{/}{%
}{/}3544548\hspace{.1pt}\discretionary{.}{%
}{.}\hspace{.4pt}3581182}}}


\bibitem{salmanowitz_impact_2018}
N.~Salmanowitz.
\newblock The impact of virtual reality on implicit racial bias and mock legal decisions.
\newblock {\em Journal of Law and the Biosciences}, 5(1):174--203, May 2018. \href{https://doi.org/10.1093/jlb/lsy005}
{doi: {{%
10\hspace{.1pt}\discretionary{.}{%
}{.}\hspace{.4pt}1093\discretionary{/}{%
}{/}jlb\discretionary{/}{%
}{/}lsy005}}}


\bibitem{schulze_2019}
S.~Schulze, T.~Pence, N.~Irvine, and C.~Guinn.
\newblock The {Effects} of {Embodiment} in {Virtual} {Reality} on {Implicit} {Gender} {Bias}.
\newblock In J.~Y. Chen and G.~Fragomeni, eds., {\em Virtual, {Augmented} and {Mixed} {Reality}. {Multimodal} {Interaction}}, pp. 361--374. Springer International Publishing, Cham, 2019.

\bibitem{schwind2017}
V.~Schwind, P.~Knierim, C.~Tasci, P.~Franczak, N.~Haas, and N.~Henze.
\newblock "{{These}} are not my hands!": {{Effect}} of {{Gender}} on the {{Perception}} of {{Avatar Hands}} in {{Virtual Reality}}.
\newblock In {\em Proceedings of the 2017 {{CHI Conference}} on {{Human Factors}} in {{Computing Systems}}}, pp. 1577--1582. {ACM}, {Denver Colorado USA}, May 2017. \href{https://doi.org/10.1145/3025453.3025602}
{doi: {{%
10\hspace{.1pt}\discretionary{.}{%
}{.}\hspace{.4pt}1145\discretionary{/}{%
}{/}3025453\hspace{.1pt}\discretionary{.}{%
}{.}\hspace{.4pt}3025602}}}


\bibitem{seitz2020}
K.~R. Seitz, J.~J. Good, and T.~C. Peck.
\newblock Shooter {Bias} in {Virtual} {Reality}: {The} {Effect} of {Avatar} {Race} and {Socioeconomic} {Status} on {Shooting} {Decisions}.
\newblock In {\em 2020 {IEEE} {Conference} on {Virtual} {Reality} and {3D} {User} {Interfaces} {Abstracts} and {Workshops} ({VRW})}, pp. 606--607. IEEE, Atlanta, GA, USA, Mar. 2020. \href{https://doi.org/10.1109/VRW50115.2020.00154}
{doi: {{%
10\hspace{.1pt}\discretionary{.}{%
}{.}\hspace{.4pt}1109\discretionary{/}{%
}{/}VRW50115\hspace{.1pt}\discretionary{.}{%
}{.}\hspace{.4pt}2020\hspace{.1pt}\discretionary{.}{%
}{.}\hspace{.4pt}00154}}}


\bibitem{spanlang2014}
B.~Spanlang, J.-M. Normand, D.~Borland, K.~Kilteni, E.~Giannopoulos, A.~Pomés, M.~González-Franco, D.~Perez-Marcos, J.~Arroyo-Palacios, X.~N. Muncunill, and M.~Slater.
\newblock How to build an embodiment lab: Achieving body representation illusions in virtual reality.
\newblock {\em Frontiers in Robotics and AI}, 1, 2014. \href{https://doi.org/10.3389/frobt.2014.00009}
{doi: {{%
10\hspace{.1pt}\discretionary{.}{%
}{.}\hspace{.4pt}3389\discretionary{/}{%
}{/}frobt\hspace{.1pt}\discretionary{.}{%
}{.}\hspace{.4pt}2014\hspace{.1pt}\discretionary{.}{%
}{.}\hspace{.4pt}00009}}}


\bibitem{tsakiris2006}
M.~Tsakiris, G.~Prabhu, and P.~Haggard.
\newblock Having a body versus moving your body: {How} agency structures body-ownership.
\newblock {\em Consciousness and Cognition}, 15(2):423--432, 2006. \href{https://doi.org/10.1016/j.concog.2005.09.004}
{doi: {{%
10\hspace{.1pt}\discretionary{.}{%
}{.}\hspace{.4pt}1016\discretionary{/}{%
}{/}j\hspace{.1pt}\discretionary{.}{%
}{.}\hspace{.4pt}concog\hspace{.1pt}\discretionary{.}{%
}{.}\hspace{.4pt}2005\hspace{.1pt}\discretionary{.}{%
}{.}\hspace{.4pt}09\hspace{.1pt}\discretionary{.}{%
}{.}\hspace{.4pt}004}}}


\bibitem{waltemate2018}
T.~Waltemate, D.~Gall, D.~Roth, M.~Botsch, and M.~E. Latoschik.
\newblock The {Impact} of {Avatar} {Personalization} and {Immersion} on {Virtual} {Body} {Ownership}, {Presence}, and {Emotional} {Response}.
\newblock {\em IEEE Transactions on Visualization and Computer Graphics}, 24(4):1643--1652, Apr. 2018. \href{https://doi.org/10.1109/TVCG.2018.2794629}
{doi: {{%
10\hspace{.1pt}\discretionary{.}{%
}{.}\hspace{.4pt}1109\discretionary{/}{%
}{/}TVCG\hspace{.1pt}\discretionary{.}{%
}{.}\hspace{.4pt}2018\hspace{.1pt}\discretionary{.}{%
}{.}\hspace{.4pt}2794629}}}


\bibitem{waltemate_impact_2018}
T.~Waltemate, D.~Gall, D.~Roth, M.~Botsch, and M.~E. Latoschik.
\newblock The {Impact} of {Avatar} {Personalization} and {Immersion} on {Virtual} {Body} {Ownership}, {Presence}, and {Emotional} {Response}.
\newblock {\em IEEE Transactions on Visualization and Computer Graphics}, 24(4):1643--1652, Apr. 2018. \href{https://doi.org/10.1109/TVCG.2018.2794629}
{doi: {{%
10\hspace{.1pt}\discretionary{.}{%
}{.}\hspace{.4pt}1109\discretionary{/}{%
}{/}TVCG\hspace{.1pt}\discretionary{.}{%
}{.}\hspace{.4pt}2018\hspace{.1pt}\discretionary{.}{%
}{.}\hspace{.4pt}2794629}}}


\bibitem{Wobbrock2011}
J.~O. Wobbrock, L.~Findlater, D.~Gergle, and J.~J. Higgins.
\newblock {The aligned rank transform for nonparametric factorial analyses using only ANOVA procedures}.
\newblock In {\em Proceedings of the ACM Conference on Human Factors in Computing Systems (CHI '11)}, pp. 143--146, 2011.

\bibitem{wu2022}
L.~Wu and K.~B. Chen.
\newblock Examining the {Effects} of {Gender} {Transfer} in {Virtual} {Reality} on {Implicit} {Gender} {Bias}.
\newblock {\em Human Factors: The Journal of the Human Factors and Ergonomics Society}, p. 001872082211452, Dec. 2022. \href{https://doi.org/10.1177/00187208221145264}
{doi: {{%
10\hspace{.1pt}\discretionary{.}{%
}{.}\hspace{.4pt}1177\discretionary{/}{%
}{/}00187208221145264}}}


\end{thebibliography}

\end{document}